\documentclass{ws-rv961x669}
\usepackage{ws-rv-van}     
\usepackage{ws-rv-thm}     
\usepackage{subfigure}     
\makeindex

\def\be{\begin{equation}}
\def\ee{\end{equation}}
\def\bea{\begin{eqnarray}}
\def\eea{\end{eqnarray}}
\def\bfq{\mathbf{q}}
\def\bfp{\mathbf{p}}
\def\bfA{\mathbf{A}}
\def\bfa{\mathbf{a}}
\def\bfr{\mathbf{r}}
\def\bfD{\mathbf{D}}

\def\bfe{\mathbf{e}}
\def\bfj{\mathbf{j}}

\def\cf{{\mbox{\tiny{cf}}}}

\def\cs{{\mbox{\tiny{CS}}}}
\def\df{{\mbox{\tiny{DF}}}}

\newcommand{\ex}[1]{\left \langle #1 \right \rangle}
\newcommand{\sex}[1]{\langle #1 \rangle}

\begin{document}

\chapter[The Half-Full Landau Level]{The Half-Full Landau Level}

\author[B. I. Halperin]{Bertrand I. Halperin }

\address{Physics Department, Harvard University\\
17 Oxford Street, Cambridge, MA 02138, USA \\
halperin@physics.harvard.edu }

\begin{abstract}
At even-denominator Landau level filling fractions, such as $\nu=1/2$, the ground state, in most cases, has  no energy gap, and there is no quantized plateau in the Hall conductance. Nevertheless, the states exhibit  non-trivial low-energy phenomena. Open questions concerning the proper description of these systems have attracted renewed attention during the last few years.  Issues at $\nu=1/2$ include consequences of particle-hole symmetry, which should be present for a spin-aligned system  in the limit where one can neglect mixing between Landau levels. Other issues concern questions of anisotropy and geometry,   properties at non-zero temperature, and effects of relatively strong disorder.  In cases where one does  find a gapped even-denominator quantized Hall state, such as $\nu=5/2$ in GaAs structures,  major  questions have arisen about the nature of the quantum state, which will be discussed briefly in this chapter.   The chapter will also discuss phenomena that can occur in a two-component system near half filling, {\it{i.e.}}, when  the total filling factor  $\nu_{\rm{tot}} $ is close to  1. 
 
  \end{abstract}


\body

\tableofcontents

\section{Introduction}

The fractional quantized Hall effect was first observed in 1982, in a two-dimenisional electron gas confined in  a GaAs heterostructure. \cite{tsui82}.  Over the next few years, many fractional quantized states were  observed in GaAs structures, almost exclusively at fractions with odd denominator. In  1987, Willett {\it{et al.}} reported the existence of a quantized state at even denominator filling fractions $\nu=5/2$ and $\nu=7/2$,\cite{willetteven87} and in 1992, Suen {\it{et al.}} \cite{Suen1992} reported observation of a plateau at $\nu=1/2$ in a wide quantum well, with (most likely) several occupied subbands. 
However, no quantized Hall state was observed at $\nu=1/2$ in a narrow GaAs quantum well  or heterostructure.  

In the vicinity of $\nu=1/2$, the Hall conductance was found to vary smoothly, essentially linearly with the Landau level filling fraction, as in a classical Hall conductor, while the longitudinal resistivity varied continuously, with a magnitude  depending on the quality of the sample. Nevertheless, after  Willett {\it{at al.}} \cite{willett90} reported  anomalous behavior in the propagation of surface acoustic waves  near $\nu=1/2$ in a high-quality  GaAs sample, in 1990, it  became clear that something non-trivial was happening in this regime.

An explanation for the surface-acoustic-wave anomaly was provided by the theory of Halperin, Lee, and Read (HLR).\cite{hlr}  In this picture,  the state at $\nu=1/2$ was described as a  Fermi sea of composite fermions, interacting with a Chern-Simons gauge field, such that the effective magnetic field felt by the fermions is zero, on average. The HLR theory also made numerous other predictions for the low-energy properties of a quantum Hall system near $\nu=1/2$, and near other even-denominator fractions, which have been verified in experiments, at varying degrees of accuracy. 

Despite its successes, the HLR theory left many questions unanswered. As in the conventional Fermi liquid theory of electrons in zero magnetic field, the HLR theory is a low-energy effective theory, which depends on parameters whose values can only be obtained from independent calculations or observations.  Moreover, the HLR theory contains infra-red divergences, whose consequences are only partially understood.  Other questions concern issues of compatibility with the requirements of particle-hole symmetry, which should be present in the limit  where the bare electron mass is taken to zero and mixing between Landau levels can be ignored, assuming that the spins of electrons in the partially full Landau level are completely aligned by the applied magnetic field.\cite{girvin84} 

Issues related to particle-hole symmetry have received considerable attention in the last few years. In particular, in 2015, D. T. Son proposed a new  description of the half-filled Landau level, employing a Fermi sea of relativistic Dirac fermions, with a manifest particle-hole symmetry.\cite{sonphcfl,son18}  While the Son-Dirac theory has several advantages over the HLR theory, it turns out that the two theories make identical predictions for physically observable quantities in many important cases.  (HLR also has some advantages over the Son theory,  in that one can more readily see how it may be derived from the microscopic  Hamiltonian for electrons in a semiconductor.)  Nevertheless, there are certain properties for which  it is not currently clear how a correct particle-hole-symmetric prediction can emerge in the HLR theory in the case where the microscopic Hamiltonian is particle-hole symmetric, so there remain open questions about whether the two-theories are fundamentally equivalent.  These issues will be a major focus of the present chapter. [See, especially, Sections \ref{formulations}, \ref{predictions},  and \ref{open}.]  Several alternate approaches to problems of the half-full Landau level and how these approaches may relate to the HLR and Son-Dirac theories, will be discussed in  Section~\ref{approaches}.  Other related issues, including possible effects of anisotropy and geometry, strong disorder, and finite temperatures, will be reviewed in Section~ \ref{issues}. 

Questions related to particle-hole symmetry also play a role in understanding the nature  of the fractional quantized Hall state that has been observed at filling fraction 5/2. We shall touch only briefly on these issues, in Section~\ref{gapped} of the present chapter, as a more detailed discussion may be found in the chapter by Heiblum and Feldman.

Thus far,  we have  implicitly assumed that the partially filled Landau level under consideration is completely spin polarized, and that there are no low-lying transverse modes or valley degeneracies  to worry about, so  we can neglect any  degrees of freedom for the electrons other than their orbital motion in the x-y plane.  It is only under this assumption that the situation at $\nu=1/2$ or $\nu=5/2$ can be described as having a half-filled Landau level equivalent to itself under a particle-hole transformation.  By contrast, if the electron spins are not completely polarized, then particle hole-symmetry, in the absence of  mixing between Landau levels, will only relate a state at filling factor $\nu$ to a state at filling $2-\nu$.   In this case, the condition of half-filling actually occurs at filling factor $\nu=1$. 

The picture of electrons with two ``internal'' components is also relevant to bilayer systems, when the active electrons are fully spin polarized. Introducing a pseudospin  index $\tau = \pm1$ to differentiate the layers, one finds that $\tau$ plays a role analogous to the  z-component of spin in a single layer system. Experimentally, the ability to make measurements with separate contacts to the two layers and the ability to vary the distance between layers relative to the magnetic length have enabled the study of a rich variety of phenomena that could not be studied in single layer systems.

For a spin-polarized bilayer, in the  absence of Landau level mixing,  the condition of an inert filled Landau level is  achieved when the total filling $\nu_{\rm{tot}}$ is equal to 2.  Thus, the half-filled condition is achieved at $\nu_{\rm{tot}}=1$, and there should be  at least an  approximate particle-hole symmetry about that value of $\nu_{\rm{tot}}$.  States of two-component systems near $\nu_{\rm{tot}}=1$ will be discussed Section \ref{bilayers} below.

\section{$\nu \approx 1/2$: HLR and Son-Dirac Formulations} \label{formulations}

\subsection{Definition of the Problem \label{sec:problem}}

We start by considering the simplest model for a  two-dimensional  system of spin-polarized interacting electrons in a strong magnetic field, with a Landau level filling fraction $\nu$ that is equal to or close to $\nu=1/2$.  We assume that in the absence of the magnetic field, the electrons can be described by a parabolic dispersion with a bare band mass $m$, so that the Hamiltonian in the field may be written in the form
\be
\label{start}
H_0 = \sum_j \frac{|\bfp_j + \bfA (\bfr_j)|^2 }{2m  } +V_2,
\ee
where $V_2$ is a two-body interaction of the form
\be
\label{twobody}
V_2 =   \frac{1}{2} \sum_{i \neq j} v_2( \bfr_i - \bfr_j).
\ee
Here $\bfr_j$ and $\bfp_j$ are the position and momentum of electron $j$,  and $\bfA$ is the vector potential due to a uniform magnetic field $B$ in the $z$-direction.
In the case where $v_2$ is a long-range potential, the Hamiltonian must include interactions between the electrons and a uniform neutralizing background,  which can be taken into account by omitting the zero-wave-vector component of $v_2$.
In the presence of impurities, one must add   a one-body potential $V_1(\bfr_j)$ which depends on position; for the present, however, we shall consider a system without impurities, so we take $V_1 = 0$.
Except where otherwise stated, we use units where the electron charge is $e = -1$, and $\hbar=c=1$. In these units, the quantum of conductance $e^2/h$ equals $1/2 \pi$, and the quantum of flux is $h/|e| = 2 \pi$. 

\subsection{Particle-Hole Symmetry}

For the system described by  (\ref{start}) and (\ref{twobody}),
 in the limit where $m \to 0 $ while the strength of the electron-electron interactions is held constant, there will be no mixing between Landau levels. This is  because the energies separating Landau levels become infinite, while the energy differences between many-body states within a Landau level remain finite.  As stated above, it can then be shown, for the case of pure two-body interactions, that there is an exact particle-hole symmetry, such that for any eigenstate of the Hamiltonian at a filling factor $\nu$, there is a corresponding eigenstate at filling factor $1-\nu$. \cite{girvin84} (This assumes that we compare systems at the same magnetic field but different electron densities.)   Furthermore, the energy difference between a pair of states at filling $\nu$ will be identical to the energy separation between the corresponding states at $1-\nu$.   At $\nu = 1/2$, this means that an energy eigenstate must either be equivalent to itself under a particle-hole transformation, or the state must be degenerate. In addition, particle-hole symmetry imposes relations between response functions and transport properties for systems at conjugate filling factors, and it imposes certain restrictions on these quantities at $\nu=1/2$, as will be discussed further below.  In the following discussions, we shall generally abbreviate particle-hole symmetry as PH symmetry.
 
In actual experiments, since $m \neq 0$, the PH symmetry should not be exact. However, in most cases the effects of Landau level mixing are small, and it should be a good approximation to neglect them.  Since numerical calculations are greatly simplified if one can neglect Landau level mixing, almost all calculations have been  carried out under this assumption. Moreover, from a theoretical point of view, the limit of $m \to 0$ is important to understand, and any complete theory should be able to describe properly this limit.  

In comparing theory with experiment, one should be aware that in experiments the filling factor $\nu$ is most frequently varied by changing the magnetic field at fixed electron density, rather than by varying the density.  In this case, the data should not be precisely symmetric  between states at $\nu$ and $1-\nu$, even if Landau level mixing is neglected. Thus, in order to explore questions of PH symmetry,  the experimental data should be corrected to  account for the change in magnetic field.

It should be noted that if one includes three-body interactions in addition to the two-body interactions, PH symmetry will be absent even if one neglects Landau level mixing.  Similarly,  one may understand effects of Landau-level mixing  by noting that in the case of small but finite $m$, three-body interactions will be generated if one uses perturbation theory to eliminate effects of higher Landau levels.   

In the presence of impurities, PH symmetry will only be strictly present if, statistically, there is an equivalence between positive and negative impurities.   However, PH symmetry may be a good approximation if the contributions of individual impurities are weak, so that  the disorder potential is well approximated by Gaussian fluctuations  about its mean.

  In this chapter, except where otherwise stated, we shall be concerned with properties of a large flat system  in the thermodynamic limit, far from any boundaries. PH symmetry will always be strongly broken at the boundaries of a sample.  In addition, PH symmetry must be defined with care on a closed curved surface without boundaries, such as a sphere.

For a system where the kinetic energy and two-body-interactions obey circular symmetry, if one can neglect Landau level mixing, then the two-body interaction at a fixed magnetic field is completely characterized by a discrete set of parameters $v_l$, the Haldane pseudopotentials, which are the matrix elements of $V_2$ between two-body states with  relative angular momentum $l$.  For spin-aligned electrons, only odd $l$ terms are relevant.  For a planar sample, the only difference between electrons in different Landau levels is that the values of $v_l$ will be different, for the same bare interaction $V_2$.  

The fact that the Hamiltonian has PH symmetry at half filling, when restricted to a single Landau level,  does not necessarily imply that the ground state will preserve that symmetry in the thermodynamic limit. It could happen that the PH symmetry is spontaneously broken, so that one can find two degenerate ground states,  related to each other by PH symmetry, but not symmetric in themselves. As will be discussed below, available evidence strongly suggests that there is no broken PH symmetry at half filling of the lowest Landau level   for the Hamiltonian (\ref{start}) with parameters appropriate to a narrow quantum well in GaAs, i.e., at $\nu=1/2$ or 3/2.   However, numerical calculations have suggested that PH symmetry should be spontaneously broken  in the second Landau level at half filling, i.e., $\nu = 5/2$ or 7/2, where a gapped FQH state is observed experimentally.  This will  be discussed, briefly, in Section \ref{gapped}  below.

\subsection{The HLR hypothesis}

The fermion-Chern-Simons approach employed in HLR began with an exact unitary transformation, a singular gauge transformation, where the many-body electron wave function is multiplied by a phase factor that  depends on the positions of all the electrons, such that the transformed Hamiltonian acquires a Chern-Simons gauge field $a_\mu$, with two negative  flux quanta attached to every electron.  The transformed problem may be expressed in Lagrangian form by the following Lagrangian density:
\be
\label{Lhlr}
\mathcal{L}_0  =  \bar{\psi}  \left( i D_t - \mu  +\frac {\bfD \cdot \bfD }{2m}
\right) \psi  -\frac {ada} {8 \pi }  + \mathcal{L}_{\rm {int}}
\ee
\be
a da \equiv  \varepsilon^{\mu \nu \lambda} a_\mu \partial_\nu a_\lambda
\ee
\be
D_\mu \equiv \partial_\mu + i \, (a_\mu - A_\mu) ,
\ee
and $\mu$ takes on the values $(0, x,y)$. 
Taking the variation of the Lagrangian with respect to $a_0$, we obtain the constraint
\be
\label{bcs}
\nabla \times \bfa = - 4 \pi  \, \bar{\psi} \psi = - 4 \pi \, n_{\rm{el}}(\bfr).
\ee
In these equations,   $\psi$ is the Grassmann field  for a set of  transformed ``composite fermions" (CFs), whose density $\bar{\psi} \psi$ is identical to the electron density $ n_{\rm{el}}(\bfr)$.  

At this stage, we have merely transformed one insoluble problem to another.  However, the transformed problem admits a sensible mean-field approximation, whereas the original problem did not.  In particular, if the Landau level is half full, so that there is one electron for each quantum of electromagnetic flux, the mean field problem describes a set of non-interacting fermions in zero magnetic field.  To go beyond mean-field theory, one must include the effects of fluctuations in the gauge field and fluctuations in the two-body potential. The central hypothesis of HLR is that, in principle,  one could obtain the correct properties of the system by  starting from the mean field solution, treating the omitted fluctuation terms via  perturbation theory. In other words, one could start with the mean field model of non-interacting fermions in zero field, and one could imagine turning on gradually the Coulomb interaction and the interactions via the gauge field, while simultaneously turning on the external magnetic field in such a manner that the effective magnetic field remains zero. 
HLR assumes that the interacting ground state can be reached from the mean-field solution by turning on the perturbing terms adiabatically, without encountering any phase transition.  Among the consequences of this assumption are that the ground state at $\nu = 1/2$ should be compressible, that there should be no energy gap, and that there should be something like a Fermi surface, with a well-defined Fermi wave vector, $k_F = 4 \pi n_{\rm{el}}$\cite{hlr,altshuler94,kim94,sternSH95}. 
This may be contrasted with the gapped  quantized Hall states at the Jain fractions $\nu = p/(2np+1)$, discussed  in  earlier work using composite fermions and a Chern-Simons gauge field. 
\cite{jaincf,lopez91,greiter92,rejaei92}. The HLR analysis was, of course, motivated by this earlier work. (See also Ref.~\refcite{kalmeyer92}.)

In analogy with the Landau theory of a Fermi system in the absence of a magnetic field, one may hope that by integrating out high frequency fluctuations of the fermion and gauge fields, one may obtain a renormalized effective theory, with some resemblance to the original theory, which can be used to calculate response functions and other phenomena at low frequencies.  The simplest guess for the low energy theory is given by a Lagrangian of the same form as the bare HLR theory, but with a renormalized effective mass  $m^*$ for the composite fermions, which one might then treat in the Random Phase Approximation (RPA).  (A renormalized mass is necessary, for example, to obtain the correct specific heat at low temperatures, or to obtain the correct energy gaps for FQH states near to $\nu=1/2$, since the energy scale is actually determined by the strength of the Coulomb interaction, rather than by the bare electron mass $m$,  when $m \to 0$. ). The meaning of the RPA , here, is that response functions are calculated by assuming the composite fermions behave as free fermions 
of mass $m^*$ in response to the effective electromagnetic field, which is the difference of the real applied electromagnetic field and the Chern-Simons fields produced by the induced variations in fermion density and current.  Specifically, by taking variations of the action governed by  (\ref{Lhlr}), one finds that a non-zero value of the fermion current density  $\bfj$ leads to a Chern-Simons electric field of the form 
\be
\bfe = 4 \pi  \hat{z} \times \bfj ,
\ee
while variations in the fermion density lead to variations in the Chern-Simons magnetic field $b = \nabla \times \bfa$ prescribed by (\ref{bcs}).  In the HLR formulation,  the density and currents of the composite fermions are the same as those of the electrons, at least in the long-wavelength limit, even after renormalization. This leads to  relations between the electro- magnetic response functions and free-fermion response functions, at the RPA level, which can be summarized as 
\be
\label{cfel}
\hat {\rho} (\bfq, \omega) = \hat {\rho}^{\cf} (\bfq, \omega) + \hat{\rho}^\cs ,
\ee
where $\hat {\rho} (\bfq, \omega)$ is the electrical resistivity tensor at  wave vector $\bfq$ and frequency $\omega$, while $\hat {\rho}^{\cf} (\bfq, \omega)$ is the corresponding tensor for free composite fermions, 
and 
\be
\label{rhocs}
\hat{\rho}^\cs \equiv  4 \pi \hat{\epsilon}
\ee
is the Chern-Simons resistivity, with  $\hat{\epsilon}$ being the unit antisymmetric tensor in two-dimensions. 
 
 If one simply uses the RPA with a renormalized effective mass to calculate the long-wavelength response at finite frequencies, one will not obtain correct answers.  For example, the response function will violate Kohn's theorem, which says that for the system defined by Eq. (\ref{start}),   the density response function at long wavelengths  should be dominated by a single pole at the bare cyclotron frequency, $\omega_c = B/m$, whereas the RPA would predict a pole at a frequency $\omega^*_c =B/m^*$.  As in conventional Fermi liquid theory, this problem is corrected if one includes effects of a Landau interaction parameter $F_1$, which describes the interaction energy associated with a uniform displacement of  Fermi surface.  An approximation which includes effects of $F_1$ together with a renormalized $m^*$ has been called the Modified Random Phase Approximation, or MRPA.  A further generalization, designed to take into account effects of the Zeeman and cyclotron energy variations in the case of a non-uniform magnetic field, has been denoted the Magnetized Modified Random Phase Approximation, or MMRPA.  However in the case of a uniform magnetic field, or in the case where the effective electron  $g$-factor is equal to two, there is no difference between MMRPA and MRPA. 
 
 A more complete theory, capable of describing arbitrary distortions of the composite fermion Fermi surface should include interaction parameters $F_l$ describing distortions of the Fermi surface for an arbitrary circular harmonic $l$.  Moreover, in order to describe response functions at non-zero wave vectors, we expect that the RPA must be further modified to include vertex corrections that depend on wave vector and frequency.


\subsection{Son-Dirac}


In the 2015 article mentioned above,\cite{sonphcfl}   Son proposed a model in which massless Dirac particles interact with a gauge field   by a Lagrangian density of the form
\begin{eqnarray}
\label{dirac}
\mathcal{L}_D = &\psi^\dagger (i D_t - \mu - i v_D \, {\bf{D}} \times {\bf{\sigma}}) \psi  +
\nonumber  \\
&+ \left[ \frac {AdA} {8 \pi }   +  \frac{adA}{4 \pi }   \right]  +
 \mathcal{L}_{\rm {int}} ,
\end{eqnarray}
\be
D_\mu \equiv \partial_\mu + i \, a_\mu ,
\ee
where  $\psi$ is a two-component Grassmann spinor, $\bfA$ is the external  magnetic  field, $\mathbf{\sigma}$ are the Pauli spin matrices, and $ \mathcal{L}_{\rm {int}} $ is a term which represents the two-body interaction $V_2$.  The velocity $v_D$ is an input parameter, like the effective mass $m^*$ in the HLR theory, which must be taken either from experiment or from an independent microscopic calculation.

In the Son-Dirac formulation, the composite fermions see an effective magnetic field  $b(\bfr)$ which is related to the electron density and the applied magnetic field in the same way as in HLR:
\be
b =  \nabla \times \bfa =  4 \pi n_{\rm{el}} - \nabla \times \bfA .
\ee
However, the electron density and the composite fermion density are not necessarily identical.  Rather, the density of Dirac composite fermions  is tied to the (local) value of the magnetic field
\be
n_\df = : \psi^\dagger \psi: =   \frac {1}{4 \pi} \nabla \times \bfA ,
\ee
which is equal to the density of electrons in the ground state at $\nu=1/2$, but may deviate from the electron density more generally.  
Similarly, the current of the Dirac fermions is related to the  local electric field  by
\be
\bfj_\df =  - \frac {1}{4 \pi} \hat{z} \times  \bf{E},
\ee
while the  effective electric field felt by the Dirac fermions is given by
\be
\bfe_\df = -\nabla a_0 - \partial_t \bfa =  \hat{z} \times (4 \pi  \bfj_{\rm{el}}) - \bf{E}   .
\ee
The electrical conductivity tensor, for a long-wavelength electric field is then given by
\be
\label{sigrhodf}
\hat{\sigma} = -  \hat{\varepsilon} \, \hat{\rho}^\df  \, \hat{\varepsilon} + \hat{\sigma}^\cs ,
\ee
where $\hat{\rho}^\df  = (\hat{\sigma}^\df)^{-1} $ is the resistivity tensor of the Dirac fermions, and
\be
\hat{\sigma}^\cs = - \frac{1}{4 \pi} \hat{\varepsilon} .
\ee

Because there is no Chern-Simons term of the form $ada$, and because the density of fermions is unchanged if the density of electrons is varied while the magnetic field is held constant, the Son-Dirac theory is explicitly particle-hole symmetric about $\nu=1/2$.  However, the Son-Dirac Lagrangian in its original form, is not sufficient to describe arbitrary deviations from the $\nu=1/2$ ground state.  As in the HLR approach, one must include the equivalent of Landau interaction parameters to describe arbitrary deviations of the shape of the Fermi surface, and one must include  vertex corrections and higher-order interactions  in order to describe correctly responses at finite wave vectors or non-linear behavior. 

\subsection{Contrasts}

There are several striking differences between the Son-Dirac and HLR formalisms.  Most obvious is the use of  two-component massless Dirac fermions in one case and non-relativistic spinless fermions in the second.  However, because there is a non-zero fermion density in both cases, and we are concerned with low energy properties of the system, neither the filled negative energy states in the Dirac model, nor  the parabolic dispersion in the case of HLR,  is of obvious importance.  The difference between the two band structures that would seem to have most effect is the presence in the Dirac case, but not in HLR, of  a Berry phase of $\pi$, as one moves adiabatically around the Fermi surface.

A second apparent difference between the two theories arises from the absence of a Chern-Simons $ada$ term in the Son formulation, and the related result that the density of fermions will be different in the two theories when the filling factor deviates from 1/2.  However, this difference may be more a matter of representation than of physics.  The density of fermions in a relativistic theory is normally defined by including only the positive energy states, and ignoring the filled sea of negative energy states.  For $\nu \neq 1/2$, the effective magnetic field felt by the fermions will be non-zero, and in the Dirac formulation there will  be  a finite density of occupied states precisely at zero energy. In the Son-Dirac formulation, one counts in the fermion density precisely half of the fermions in zero energy states.  If, instead, one were to omit all zero-energy  fermions, the density of fermions would  then be equal to that of the electrons, just as in the HLR approach. 

The last point can be made clearer if one employs  a generalization of the Son approach, which enables one to consider Dirac particles with a non-zero mass. \cite{PSV}   Consider the Lagrangian density 
\begin{eqnarray}
\label{mdirac}
\mathcal{L}_D = &\psi^\dagger (i D_t - \mu - i v_D \, {\bf{D}} \times {\bf{\sigma}} - m_D \, \sigma^z) \psi  +
\nonumber  \\
&+ \left[ \frac {AdA} {8 \pi }   +  \frac{adA}{4 \pi } - \frac {ada} {8 \pi } \frac{m_D}{|m_D|}  \right]  +
 \mathcal{L}_{\rm {int}} ,
\end{eqnarray}
\be
D_\mu \equiv \partial_\mu + i \, a_\mu ,
\ee
  We are interested in a situation in which the Fermi level is inside the band of positive energy fermion states 
 and the lower Dirac band has been integrated out, which is the origin of  the  $ \frac{1}{8\pi}ada \, {\rm{sign}}  (m_D) $ term in (\ref{mdirac}). 

The  Lagrangian (\ref{mdirac}) reduces to (\ref{dirac})  if one takes the limit $m_D \to 0$. In this limit, the contribution of the $ada$ term in (\ref{mdirac}) in any application  is  precisely canceled by  the contribution from the Berry curvature, which is completely concentrated  at the bottom of the occupied states in the positive energy Dirac band.  (By convention, this contribution, which is ill-defined for $m_D=0$, is omitted in that case.)    Thus, in the limit  $m_D \to 0$,  the Lagrangian (\ref{mdirac}) may be replaced by the form (\ref{dirac}),    in which $m_D$ is precisely zero and the $ada$ term is simply omitted; {\it i.e.}, there is no longer a Chern-Simons term in the action for the gauge field $a_\mu$.   In the following discussion, we confine ourselves to the case $m_D$=0, except where otherwise specified.

The question we will return to below is to what extent the differences in formulation between Son-Dirac theory and HLR result in differences in predictions for physical observables, and whether in the end the two theories are fundamentally different.

\subsection{Infrared divergences} \label{divergences}
Discussions of the asymptotic low-frequency behavior at $\nu=1/2$ are complicated by the occurrence of infrared divergences, predicted by both the HLR and Son-Dirac theories. In the case of Coulomb-like electron-electron interactions, which  fall off as $1/r$ at large electron separations $r$,  it was predicted in Ref.~\refcite{hlr}  that there  will be a relatively-innocuous logarithmic 
divergence in the composite-fermion effective mass.  Also the decay rate of a composite fermion near the Fermi energy should be smaller, by a logarithmic factor than the energy. Subsequent calculations using a renormalization group concluded that the mass would only diverge as a the square-root of the logarithm of the distance from the Fermi surface. \cite{NayakWilczek94}  In either case,  the composite fermion system may be considered  a kind of marginal fermi liquid,  interacting with a gauge field at long wavelengths.   Since the infrared divergences are expected to be identical in the Son-Dirac and  HLR theories, it seems reasonable to ignore the divergence in $m^*$ when comparing the two theories.  Alternatively, the infrared divergences can be eliminated entirely if one wishes to consider a model where the interaction $v_2$ falls off more slowly than $1/r$.

In the case of short range interaction, the infrared divergences are much stronger, and quasiparticles can no longer be precisely defined.  Nonetheless, important features of the fermi surface are expected to survives, such as the existence of some type of singularity in various response functions at wave vector $q= 2k_F$, at $\nu=1/2$. Moreover, many of the quantities discussed below, are predicted in both Son-Dirac and HLR to have a  behavior independent of the value of $m^*$, and it is expected that these predictions should remain correct even in the case of short-range interactions. Other quantities, such as the compressibility, which for a non-interacting Fermi system would be proportional to the value of the effective mass, are found to remain finite, essentially  because the divergence in $m^*$ is canceled by  similar divergence in the various Landau parameters $F_l$. 
\cite{kim94,altshuler94,sternSH95}

\section{Predictions for Behavior near $\nu=1/2$ } \label{predictions}

\subsection{Predictions where HLR and Son-Dirac agree} \label{3.1} 

\subsubsection*{Nearby quantized states.} 

As mentioned above, there is large set of phenomena where predictions of the Son-Dirac theory coincide at the RPA level  with the previous predictions of HLR.  In both cases, at $\nu=1/2$ the density of composite fermions is equal to the density of electrons, $n_{\rm{el}}= B/4 \pi$,  and 
the mean-field ground state at $\nu=1/2$ is a sea of composite fermions  in zero effective magnetic field, with a Fermi wave vector given by
\be
k_F = [4 \pi n_{\rm{el}}]^{1/2} .
\ee
Slightly away from $\nu=1/2$, the fermions see an effective magnetic field 
\be
\Delta B = B - 4 \pi n_{\rm{el}} .
\ee
In  both theories, a fractional quantized Hall state with an energy gap is expected when the the density of composite fermions is such as to fill an integer number of Landau levels in the effective magnetic field $\Delta B$. In
HLR, as in Jain's composite fermion theory,  the density of composite fermions  is equal to $n_{\rm{el}}$,  and the gapped state  occurs when 
\be
\label{fqhhlr}
 2 \pi n_{\rm{el}} = p \Delta B ,
 \ee
with $p$ a positive or negative integer. The actual filling fraction is then
\be
\label{fqhjain}
\nu \equiv \frac {2 \pi n_{\rm{el}} }{  B}  =  \frac {p} {(2p+1)}  = \, \frac {1}{2} - \frac {1} {2(2p+1)}.
\ee

In the Son-Dirac theory, due to the Berry phase,  one predicts that the
Fermi energy is in an energy gap when the density of Dirac fermions $n_{\rm{DF}}$ satisfies
\be
\label{fqhson}
2\pi n_{\rm{DF}} = p_{\rm{DF}}  \Delta B ,
\ee
where $p_{\rm{DF}}$ is  a {\em half} odd integer.  However, $n_{\rm{DF}}=   n_{\rm{el}} /2 \nu    $,  
which differs from  the density of electrons, when $\nu \neq 1/2$.  Using simple algebra, one finds
\be
\nu = \frac{1}{2} - \frac {1} {4 p_{\rm{DF}}} .
\ee
Consequently,  the Son-Dirac theory leads to the  {\em{same}}  values of $\nu$ for the gapped states as Jain's theory or HLR [Eq.~(\ref{fqhjain})], with the identification  $p_{\rm{DF}} = p + 1/2$.    The allowed electron densities at the gapped states are  symmetric about $\nu=1/2$, if the magnetic field is held fixed.

\subsubsection*{Energy gaps.}

In the mean field theory based on HLR, the energy gap at a quantized Hall state of the form (\ref{fqhjain}) is given by 
\be
\label{gap}
 E_g = \frac {|\Delta B|} {m^*} \, ,
\ee
where $m^*$ is the renormalized effective mass at $\nu=1/2$. If this effective mass and the energy gaps are calculated at fixed magnetic field while the electron density is varied, then the predicted energy gaps will be symmetric about $\nu=1/2$, as required by PH symmetry.  

In Ref. \refcite{hlr},  HLR considered the effects  of the predicted infrared divergences in $m^*$ on the size of the energy gaps.   For the case of Coulomb interactions, where the divergence is only logarithmic, it was  argued that (\ref{gap}) could still be used, provided that $m^*$ was evaluated self-consistently at an energy distant from the Fermi surface by an amount  of order $ E_g$.   In effect this leads to an asymptotic behavior, for $\nu \to 1/2$, of the form \cite{sternSH95}
\be
\label{loggap}
 E_g  \sim \frac{\pi}{2} \frac {e^2}{\epsilon l_B} \frac{1}{ |2p+1| (C + \ln |2p+1| ) },
 \ee
where $\epsilon$ is the background dielectric constant, $l_B$ is the magnetic length, and $C$ is a constant which depends on the short range behavior of the interaction. This result is also consistant with PH symmetry.
For the case of short range interactions, where one predicts $m^* \propto |\delta E|^{-1/3}$,
one is led to an asymptotic prediction 
\be
E_g  \propto |\Delta B|^{3/2} ,
\ee 
but there is no precise prediction for the prefactors. 

As remarked earlier, however,  Nayak and Wilczek,\cite{NayakWilczek94}  have argued, using a renormalization group analysis applied to the HLR Lagrangian, that the effective mass divergences should  have a somewhat different form than those predicted by HLR in Ref. \refcite{hlr}.  In the case of Coulomb interactions, they predict that effective mass  should diverge as the square-root of the logarithm of the the distance from the Fermi surface, which would imply that the energy gaps of the Jain states
should vary asymptotically as 
\be
\label{sqrtloggap}
E_g \propto |\Delta B | / [\, \ln |\Delta B|]^{1/2} .
\ee
Although there are no published analyses of the effects of infrared divergences using the Son-Dirac theory, the predictions should be identical to those based on the  HLR Lagrangian for all of these quantities. 

Morf, d'Ambrumenil, and Das Sarma have compared the predictions of (\ref{loggap}) with results for energy gaps at fractions 1/3, 2/5, 3/7, and 4/9 that they obtained by extrapolation of exact diagonalizations on a sphere. \cite{morfMAS02} They found that (\ref{loggap}) fit their data  points better than the simple formula $E_g = C / |\Delta B|$, though both formulas contain a single fitting parameter.  However, it is impossible to draw firm conclusions from such small systems, and the data  would also be compatible with an asymptotic dependence of the form (\ref{sqrtloggap}).

\subsubsection*{Conductivity at non-zero wave vector.}

An important result of HLR was the prediction that the longitudinal wave-vector-dependent electrical conductivity at $\nu=1/2$, in the limit of vanishing frequency and small finite wave vector, should be given, in the absence of impurities, by
\be
\label{sigq}
\sigma_{xx}(q) = \frac{q}{8 \pi k_F} \, ,
\ee
where we have taken the  the wave vector $\bfq$ to lie in the x-direction. Note that this result is independent of $m^*$, and it is believed that the result should be unaffected by the infrared divergences in the energy spectrum of composite fermions.

In the presence of impurities, (\ref{sigq}) should hold for $q$ larger than the inverse of the composite fermion mean free path  $l_{\rm{cf}}$, which we assume to be long compared to the Fermi wavelength. 
For $q l_{\rm{cf}} <1$, $\sigma_{xx}$ should become independent of $q$ and should reduce to the macroscopic longitudinal conductivity, $\propto l_{\rm{cf}}^{-1}$.     The wave-vector dependence of the  conductivity $\sigma_{xx}$ is responsible for the shift in the propagation velocity of a surface acoustic wave near observed by Willett {\it {et al.}} in 1990,\cite{willett90} which was mentioned in the Introduction as evidence that something peculiar was happening at $\nu=1/2$

Within HLR, the result (\ref{sigq}) is a consequence of  the fact that the transverse wave-vector-dependent conductivity for non-interacting composite fermions, $\sigma^{\rm{cf}}_{yy}(q) $,  actually diverges as $q^{-1}$, in the absence of  impurity scattering. 
To obtain the electrical conductivity,  according to Eq.~(\ref{cfel}), one should first calculate  the CF resistivity tensor,  by  inverting  the CF conductivity tensor.  At $\nu=1/2$, in the absence of impurities,  the CF resistivity tensor is diagonal, so $\rho^{\rm{cf}}_{yy}(q) $  is just the inverse of $\sigma^{\rm{cf}}_{yy}(q) $.  In $\hat{\rho}$, however, following Eq.~(\ref{cfel}), the off-diagonal part is much larger than the diagonal,  and one obtains the result   (\ref{sigq}) for $\sigma_{xx}(q)$ on inverting $\hat{\rho}$.  

The prediction (\ref{sigq}) is also obtained  in the Son-Dirac theory, using Eq.~(\ref{sigrhodf}).   More generally, the anomalous wave-vector-dependence of the electrical conductivity reflects that fact that
 that composite fermions at $\nu=1/2$ can travel in a straight line over distances that are very large compared to the microscopic magnetic length. 

What happens to these results as one deviates slightly from $\nu=1/2$?  As the composite fermions now see an effective magnetic field $\Delta B$,  we should expect them to move in circles, with an effective cyclotron radius 
\be
\label{RckF}
R_c^* = k_F/ |\Delta B |. 
\ee
The non-local conductivity of the composite fermions will clearly be cut off for distances larger than $2R_c^*$ and $\sigma_{xx}(q)$ will accordingly deviate from its $\nu=1/2$ value when $|\Delta B|$ is large enough that  $q R_c^* <1$. However, the dependence on $|\Delta B|$ is not generally  monotonic. It was predicted in HLR that for a fixed value of $q$, there should  be {\em oscillations} as a function of $|\Delta B|$ in the region $q R_c^* >1$, provided that $l_{\rm{cf}}$ is sufficiently large. Observation of the consequent magneto-oscillations  in  surface acoustic wave propagation, by Willett  {\it{et al.}} in 1993,\cite{willett93} provided an important confirmation of the HLR theory.  

The HLR prediction for magneto-oscillations in $\sigma_{xx}(q)$ was based on a semiclassical analysis of the dc conductivity tensor for composite fermions in the presence of the effective magnetic field  $\Delta B$ and weak impurity scattering.  Roughly speaking, the features in $\sigma_{xx}(q)$ could be understood as arising from a commensurability  condition between the effective cyclotron diameter $2 R_c^*$ and the wave length $2 \pi / q$. More precisely, the analysis predicted that for small $q$ and  very small disorder, there would be peaks in $\sigma_{xx}(q)$ as a function of $\Delta B$, which would occur when 
\be
\label{semiclass}
q R_c^* = z_n ,
\ee
where $z_n \approx \pi (n + \frac{1}{4})$ is the $n$-th zero of the $J_1$ Bessel function.  Equivalently, using (\ref{RckF}), the peaks should occur at wave vectors given by
\be
\label{qn}
q_n = z_n |\Delta B| / k_F .
\ee  

Related commensurability oscillations have been predicted for other quantities near $\nu=1/2$.  We shall discuss this subject further below, since possible deviations from the predictions of (\ref{qn}) have played a role in comparisons between HLR and Son-Dirac.

\subsubsection*{Hall conductance at $\nu=1/2$}

If mixing between Landau levels is neglected, 
PH symmetry requires  that the  Hall conductivity at $\nu=1/2$,  in response to a spatially uniform electric field, should be precisely given by
\be\label{sig}
\sigma_{yx} = - \sigma_{xy} = \frac {1} {4 \pi},
\ee
regardless of the applied frequency.  This  should be true even in the presence of impurities, provided that the disorder potential $V_{\rm{imp}}$ is PH symmetric in a  statistical sense.  (This condition means that if one chooses the  uniform background potential such that the average  $\langle V_{\rm{imp}} \rangle = 0$, then all odd moments of the disorder potential must vanish.)

In 1997, Kivelson {\it{et al.}} pointed to a possible difficulty in reconciling the HLR theory with this requirement, in the context of the dc conductivity.\cite{klkgphhlr}  Their argument  went as follows.  
Within the HLR approach, the electron resistivity  tensor and the  resistivity tensor of the composite fermions are related according to Eqs.~(\ref{cfel}) and (\ref{rhocs}) above.  This means that 
in order to obtain the PH symmetric result (\ref{sig}) for $\sigma_{yx}$, when $\rho_{xx} \neq 0$, it is  then necessary that $\sigma^{\rm{\, cf}}_{xy }=  - 1 / 4 \pi.$  However, it was argued that $\sigma^{\rm{\, cf}}_{yx}$ is necessarily equal to zero at $\nu=1/2$.  This seemed evident because, in the absence of impurities, the composite fermions see an average effective magnetic field equal to zero, which is effectively invariant under time reversal. The presence of impurities leads to non-uniformities in the electron density, which lead to local fluctuations in the effective magnetic field $b(\bfr)$ that turn out to  be the dominant source of scattering of composite fermions, under conditions where the correlation length for the impurity potential is large compared to the Fermi wave length.  If the impurity potential is statistically PH symmetric, then there will be equal probability to have a positive or negative value of $b$  at any point, so that the resulting perturbation to the composite fermions should again be invariant under time reversal in a statistical sense, which would imply that  $\sigma^{\rm{\, cf}}_{yx}=0$.

However, there is a flaw in this argument. As  was shown  by Wang {\it{et al.}} in 2017,\cite{WangCoopHalpStern}   when the HLR theory is  properly evaluated, it actually  gives the correct PH symmetric result for the dc conductivity in this problem, at least to order $l_{\rm{cf}}^{-2}$.  The error in the earlier reasoning is that while the fluctuations in the electrostatic potential are highly screened at long wavelengths, and their contribution to the CF scattering rate is small compared to the fluctuations in $b(\bfr)$, they must still be taken into account when computing $\sigma^{\rm{\, cf}}_{yx}$. The residual potential fluctuations are correlated with  fluctuations in $b(\bfr)$ in a manner which breaks the statistical time-reversal symmetry present in the $b$ fluctuations alone. For example, the local density of electrons, and therefore of CFs, 
 will be slightly larger in a region with $b>0$  and smaller in a region with $b<0$. When these correlated fluctuations are taken into account, it turns out that one recovers  
precisely the result
$\sigma^{\rm{\, cf}}_{yx}= - 1 / 4 \pi$ required by PH symmetry.   Note that $\sigma^{\rm{\, cf}}_{yx}$ is actually a small correction to the leading diagonal piece, $\sigma^{\rm{\, cf}}_{xx}$, which diverges as  $l_{\rm{cf}}$ in the limit of weak disorder.

\subsubsection*{Weiss oscillations.}

An effect closely related to the above-mentioned oscillations in $\sigma_{xx}(q)$, is the oscillatory behavior of the long-wavelength dc magnetoresistance (``Weiss oscillations")  in the presence of an external periodic potential with a period $2 \pi / q$. \cite{smet98,willett99,zwerschke99,smet99,kamburov14}
  Under appropriate conditions, if the wave vector $q$ is fixed and  the electron concentration is varied at fixed magnetic field,  theory predicts that there should be a series of minima in the longitudinal resistance which should occur, at least approximately, when the conditions satisfy  (\ref{qn}).  

If one wishes to treat this equation as a precise prediction, however, one must  specify precisely the definition of the Fermi wave vector $k_F$.  If one defines $k_F$ as the square-root of $4 \pi$ times the density of composite fermions, which is equal to the density of electrons in HLR but is determined by the density of flux quanta in Son-Dirac, one would find that the wave vectors q given by  (\ref{qn}) would agree in the two theories at first order in $|\Delta B|$ but would disagree at order $|\Delta B|^2$.  Furthermore, the HLR result would  violate PH symmetry at this order.\cite{kamburov14} 

However, a more careful evaluation of the HLR theory found the same result for the positions of the resistance minima as the Son-Dirac theory, at least to order $|\Delta B|^2$, so this discrepancy is eliminated.   \cite{WangCoopHalpStern,CheungCRM16 } The analysis in both cases was carried out at the RPA level, assuming a temperature small compared to the effective Fermi energy $E_F^* = k_F^2/m^*$, but larger than $|\Delta B| / m^*$. Under these conditions,  quantum effects due to discreteness of the CF energy spectrum, such as the formation of quantized fractional Hall states,  can be ignored.  However, commensurability effects related to the diameter of the semiclassical CF cyclotron orbits can persist to higher temperatures. The calculations assume that there is a small amount of background disorder, with a large but finite CF mean-free-path, and the results are only precise in this limit.

We note that, similar to the case of $\sigma_{xx}(q)$, in order to get the correct answer here using the HLR formulation, it is necessary to take into account the variations in   the screened electrostatic potential as well as in the effective magnetic field $b$ resulting from the external periodic perturbation.  The contribution of the  electrostatic potential, though small, leads to a phase shift in the oscillations, which is just sufficient to bring the calculation based on HLR into agreement with that based on Son-Dirac. 

Weiss oscillations have been of particular interest because experimental measurements have been able to locate the induced minima in magnetoresistance with great accuracy.\cite{kamburov14}  (See the discussion in the chapter by M. Shayegan.)   Experiments have demonstrated PH symmetry for these positions with an accuracy that is able to rule out an asymmetry of the size that would have been expected from a na\"ive use of HLR. Of course, this is not a test of HLR vs Son-Dirac, since both theories actually predict the same results for the quantities in question. Interestingly, the measured positions do deviate from the predicted positions by an  amount  which is PH symmetric but is of a similar magnitude to the difference between Son-Dirac and na\"ive HLR. The sign and magnitude of the observed deviation is such that the data can be well fit by the form (\ref{qn}) with a choice of $k_F$ dictated by the density of electrons for $\nu>1/2$ and by the density of holes for $\nu < 1/2$.  This means that if one wishes to fit the data to (\ref{qn}), one should choose $k_F$ to satisfy $k_F l_B = [\min (2 \nu , 2-2\nu) ]  ^{1/2} $, rather than the value $k_F =l_B^{-1}$ predicted by the RPA-type  calculations of Refs. \refcite {WangCoopHalpStern} and \refcite{CheungCRM16}. The reason for this observed deviation 
 is not known. It may be a result  of  corrections due to factors such as to the finite density of impurities, but it could be that there would be corrections of the observed magnitude even in the limit of small impurity density.  Recent work by Mitra and Mulligan has suggested that the discrepancy may be an effect of gauge fluctuations.\cite{MitraMulligan19}

\subsubsection*{Minima in the magneto-exciton spectrum and maxima in the static structure factor.}

Another quantity that has been predicted to show commensurability oscillations near $\nu=1/2$ is the spectrum $\omega(q)$  of neutral excitations in a quantized Hall state of the form  $\nu=p/(2p+1)$, for large values of $|p|$. The lowest frequency branch is  predicted to have a series of minima, known as magnetoroton  minima, which occur at wave vectors given by (\ref{qn}) in the limit of large $|p|$ or small $|\Delta B|$.\cite{ SimonHalperin}   As in the case of the Weiss oscillations, a naive application of the HLR theory, in which $k_F$ is determined by the density of electrons, would predict a PH asymmetry at order $|\Delta B|^2$ .  However, a correct evaluation of
the formulas in Ref.~\refcite{SimonHalperin} gives results which are equivalent to setting 
$k_F = l_B^{-1}$ in (\ref{qn}). Thus one obtains
 a  PH symmetric result,  in agreement with the Son-Dirac approach.  
\cite{WangCoopHalpStern}. 

Associated with the minima of the magnetoroton spectrum are maxima in the projected static structure factor $\tilde{S}(q)$.  Specifically, if $\tilde{\rho}(\bfq)$ is the density operator at wave vector $\bfq$, projected into the lowest Landau level, then $\tilde{S}$ may be defined as 
\be
\tilde{S}(q) = \ex{ \tilde {\rho}(\bfq)  \tilde{\rho}( - \bfq)} \approx  - i \int \frac{d \omega} { 2 \pi}  
\chi (\omega + i 0^+ {\rm{sgn}} (\omega), q),
\ee
where $\chi$ is the electron-density response function, and the integral is restricted to frequencies smaller than the bare cyclotron frequency $\omega_c$. The last equality becomes exact in the limit $\omega_c \to \infty$, with interactions held fixed,  and the  integral is then taken over all finite frequencies.

Nguyen and Son \cite{NguyenSon}  have obtained  an analytic formula for  $\tilde{S}(q)$ for  the Jain states in the limit of large $|p|$, which may be written as  
\be
\label{sqan}
\tilde{S}(q) = \frac { 1 + |2p+1|} {4} (q l_B)^4 \,\frac {J_2(z) } {z J_1(z)} ,
\ee
where $z = |2p+1| q l_B$. This indeed predicts sharp maxima at the positions given by (\ref{qn}), with $k_F=l_B^{-1}$.  The quantity $|2p+1|$ is symmetric about $\nu=1/2$, as  it is  the denominator of the fraction $\nu$  for the Jain state.

Balram {\it{et al.}} \cite{balram15} have calculated the electron pair-correlation functions  for a series of Jain states close to $\nu=1/2$, using CF trial wave functions, for up to 200 particles in both spherical and torus  geometries. They extract an effective  value of $k_F$ for each filling factor by fitting the results to a functional form with oscillations controlled by $k_F$, and they plot the results  for $k_F l_B$ as a function of  $(\nu - 1/2)$.  The results exhibit PH symmetry to a very good approximation, and they extrapolate to a value close to the required value $k_F l_B = 1$ at $\nu=1/2$ . On either side of $\nu=1/2$, however, the extracted values are close to the values one would obtain by defining $k_F$ by the density of minority carriers, 
$k_F l_B = [\min (2 \nu , 2-2\nu) ]  ^{1/2} $, deviating from the predicted value $k_F l_B = 1$ in the same sense as was found in the Weiss oscillation experiments. On the other hand,  Nguyen and Son argue that the result (\ref{sqan}) should be correct even at the next to leading order in $1/|p|$, at least in the case of short-range electron-electron interaction, which would seem to prevent deviations of this form in that case. It is not known what is the significance of the difference between this prediction and the results of Balram {\it{et al.}}

Because the static structure factor and the magnetoroton spectrum can be defined in the absence of impurities, they are more convenient to investigate theoretically than the Weiss oscillations. The
exciton modes are predicted to be undamped in the vicinity of the magnetoroton minima,  so the positions of these minima can also be precisely defined in principle. However, to the best of our knowledge there are no experiments that have been able to measure either the excitation spectrum or the electron structure factor 
at small wave vectors close to $\nu=1/2$  with a precision that would be relevant here.

\subsection{Areas where HLR and Son-Dirac disagree at the RPA level} \label{disagree}

Although the positions of maxima in $\tilde{S}(q)$ and minima in the magnetoroton spectra at the Jain states  are identical in the HLR and Son-Dirac, at least to order $| \Delta B|^2$, there are deviations at this order between the two theories in their predictions for the frequency spectrum away from the minima, and for $\tilde{S}(q)$ away from its maxima.  An important example is in the behavior of $\tilde{S}(q)$ in the limit $q \to 0$.  

It is known that $\tilde{S}(q)$ must vanish proportional to  $q^4$, for $q \to 0$, when the system is in a gapped quantized Hall state. Furthermore, we may  write 
\be
\label{Stilde} 
\tilde{S} (q) = \tilde{s}_4 \, q^4  l_B^4 \,  + \, {\cal{O}} (q^6 l_B^6) ,
\ee
where $\tilde{s}_4$ is a constant.

Following the notation of Ref. \refcite{NguyenGolkar17}, we introduce a parameter $N>0$ which is equal to our parameter $p$ for $p >0$, {\it i.e.},for  $\nu < 1/2$, but $N = | p | - 1$, for $p<0$ or $\nu > 1/2$.  Thus, we may write, in the two cases,
\be
\label{nuN}
\nu =\frac {1}{2} \mp \frac {1}{2(2N+1)} .
\ee
Note that $(2N+1) = |2p+1|$, and the Jain states are obtained when $N$ is a positive integer.  
As follows from Eq.~(\ref{sqan}), the  Son-Dirac formalism is consistent with (\ref{Stilde}), with the prediction
\be
\label{s4son}
\tilde {s}_4 =  \frac {2N+1}{32} ,
\ee
for both $\nu > 1/2$ and $\nu<1/2$. This  is consistent with particle-symmetry, and in fact coincides with exact results,  obtained by other means, for electrons  confined to the lowest Landau level.  
However, as was shown in   Ref. \refcite{NguyenGolkar17},   if  $\tilde{S}(q)$ is evaluated at a Jain state in the HLR formalism at the MRPA level, with the $F_1$ parameter chosen so that the Kohn mode is at infinite frequency, one obtains the  results
\be
\tilde {s}_4 =  \frac {1}{8} \frac {N^2} {2N+1} ,  \,\,\,\, (\nu<1/2) ,
\ee
\be
\tilde {s}_4 =  \frac {1}{8} \frac {(N+1)^2} {2N+1} ,  \,\,\,\, (\nu >1/2) .
\ee
Although these results agree with (\ref{s4son}) to leading order in $N$, and are PH symmetric to that order, they deviate from the correct results and violate PH symmetry at the next order.  

Another property where HLR at the RPA level gives an incorrect answer for electrons confined to the lowest Landau level,   is for the $q^2$ correction the wave-vector-dependent Hall conductivity  at $\nu=1/2$. \cite{LevinSon}  Specifically,  we define $\sigma_H (q) \equiv - j_x(\bfq) / E_y(\bfq)$, where 
$\bfj (\bfq)$, is the electrical current at wave vector $\bfq$, induced by an electric field in the y-direction, $E_y(\bfq)$, computed    in the limit  $q \to 0, \,\, \omega \to 0, \,\, v_Fq / \omega \to 0$.  (As a result of Onsager symmetry, this quantity will be independent of the direction of $\bfq$ in a system with overall rotational symmetry.)
The exact result for this quantity, in the limit $m \to 0$, is given by
\be
\sigma_H(q) = \frac{1}{4 \pi} \left( 1 - \frac {q^2 l_B^2} { 4} \right)  +...  ,
\ee
where the omitted terms vanish faster than $q^2$. 
However, the term proportional to $q^2$ is absent in an analysis based on the HLR theory. On the other hand, the HLR formulation does  lead to a correct result for  $\sigma_H(q)$ in the gapped Jain states. Following the parametrization (\ref{nuN}), one finds 
\be
\sigma_H(q) = \frac {1}{2 \pi} \left(  \nu  \mp q^2 l_B^2 \frac{N^2}{2N+1} \right) .
\ee
The deviations from $1/4 \pi$ are  antisymmetric about $\nu=1/2$, as required.

Geraedts et al.\cite{geraedtsnum} have given us an important example where the Son-Dirac theory gives a highly non-trivial prediction that is not evident in the HLR approach, but which, nevertheless, does not seem to be incompatible with HLR.  
The authors introduce an operator $P(\bfr)$, which is proportional to $n_{\rm{el}} (\bfr)  \nabla^2 n_{\rm{el}} (\bfr)$ projected to the  lowest Landau level, and they study the correlation  function for the Fourier transform, $\langle P_{ - \bfq} P_{\bfq} \rangle$,  for $q$ close to $2 k_F$. According to the Dirac theory, this correlation function should have no observable singularity at $q=2k_F$, because $P(\bfr)$ is even under PH inversion, and fluctuations in such quantities should not give rise to backscattering across the Fermi surface at $q=2 k_F$. Geraedts et al.  studied this correlation function numerically, for electrons confined to the lowest Landau level at half filling, using density-matrix renormalization group (DMRG) methods, and  found the $2k_F$ singularity to be missing, as predicted. At the same time, they do observe a singularity at $q = 2 k_F$  in the density correlation function   $\langle  n^{\rm{el}} _{ - \bfq}  n ^{\rm{el}}_{\bfq} \rangle$, which is consistent with the Son-Dirac theory, because the electron density operator is not even under PH inversion. 

There does not seem to be any  obvious reason in HLR theory  why $\langle P_{ - \bfq} P_{\bfq} \rangle$ should be immune from a singularity at $q=2k_F$, even if one imposes the requirement of particle-hole symmetry. 
In order to actually calculate this response function in the HLR theory, however, one would have to know the correct form of the renormalized vertex that couples $P_\bfq$ to the composite fermions at $q=2 k_F$.  In general, the operator $P_\bfq$ should have a portion that couples to the CF density operator at $2k_F$ and  a portion that couples to 
multi-particle excitations. Although correlations of the CF density operator will have a $2k_F$ singularity in the HLR formalism,
if it happens that coupling of $P$  to the CF density operator vanishes at $q=2k_F$ in the limit of $m \to 0$, there will be no $2k_F$ singularity in   $\langle P_{ - \bfq} P_{\bfq} \rangle$  in this case.  However, we are not aware of any {\it a priori} argument that  this should happen.


\section{Other approaches to $\nu=1/2$}  \label{approaches} 

There are, of course, many  theoretical approaches to the behavior of electrons in a partially filled Landau level other than the HLR and Son-Dirac descriptions.  Although in many cases these approaches are not designed to describe precisely the asymptotic low-energy behavior in the limit of $\nu \to 1/2$, they may nonetheless have very good  numerical accuracy and they can often help to understand the physics in this limit.    We discuss here several of these alternate approaches.

\subsection {``Hamiltonian" Formulation and Description as a Dipolar Fermi Liquid}

As composite fermions in the Son-Dirac description do not couple directly to the electromagnetic gauge field $\bfA(\bfr, t)$,  it is possible to interpret them as neutral objects.  Indeed, Son's construction was partly inspired by works in the earlier literature proposing that the actual low-energy quasiparticles, obtained from the bare fermions of the HLR theory after ``screening" by relaxation of the high-frequency plasma modes occurring above the bare cyclotron of the electrons, are electrically neutral at $\nu=1/2$ but carry an electric dipole moment $\bf{d}$ determined by the canonical momentum $\bfp$ of the quasiparticle.\cite{read94,read96,rsgm97,read98,Pasquier1998,dhleephcf98}   Specifically, the relation between these quantities is
\be
{\bf{d}} = l_B^2 \, \hat{z} \times \bfp
\ee
Thus, the neutral  CFs of  Son-Dirac theory are, in some sense,  closer to the true low-energy quasiparticles than the charged CFs of the HLR picture.

In one notable approach,  Murthy and Shankar\cite{rsgm97}  obtained  neutral quasiparticles  starting from the unrenormalized HLR Hamiltonian, after  a unitary transformation in which the field operators for CFs and Chern-Simons vector potential were transformed to  new operators $\psi_{\rm{MS}}$ and $\bfa_{\rm{MS}}$, which  were chosen to have a number of  desirable properties. For example, the fermion operators $\psi_{\rm{MS}}$ are decoupled from the gauge fields  $\bfa_{\rm{MS}}$ in the transformed Hamiltonian in the long-wavelength limit.  The gauge fields describe the high-energy magnetoplasma modes, while the  transformed fermions, which are important for the low frequency behavior,  carry the desired electric dipole moment at $\nu=1/2$.  (A separate benefit of this approach was that  by treating the transformed Hamiltonian in a suitable approximation, Murthy and Shankar were able to obtain estimates of the quasiparticle effective mass that were tied to the electron-electron interaction and remain finite in the limit  $m \to 0$.)  In more recent work, Murthy and Shankar have generalized their Hamiltonian approach to a PH-symmetric description in terms of Dirac fermions whose density is determined by the magnetic field, as in Son's formulation.\cite{msgmp15} 

As noted by Murthy and Shankar, their formulation of the $\nu=1/2$ problem has strong analogies to the Bohm-Pines description of a three-dimensional Fermi gas with long-range $(1/r)$ Coulomb interactions.\cite{BohmPines1953}.   There, too, one employs a unitary transformation to separate low-energy fermions from a high-energy plasma mode. By contrast, the original HLR formulation is more like the Landau-Silin approach to the three-dimensional Coulomb gas,  where quasiparticles are renormalized by the short-range portions of the Coulomb interaction, but they retain the original electron charge and still interact with the long-range part of the Coulomb interaction.

In order to properly describe the low energy properties of the electron system at $\nu=1/2$, however,  the Fermi liquid formed by the dipolar quasiparticles must have some peculiar properties.  In particular, the $l=1$ Fermi liquid parameter $F_1$ must have a critical value such that a uniform displacement of the Fermi surface costs no energy.\cite{HalperinStern1998}  In fact, the dipolar Fermi liquid will have zero-energy modes even at finite wave vectors, similar to gauge degrees of freedom, which are not observable in any physical measurement. The relation between the peculiar Fermi liquid theory of dipolar quasiparticles and HLR theory was explored in some detail in  Ref.~ \refcite{sternetal99}.

The close relation between Son-Dirac CFs and the actual low energy excitations is less clear away from $\nu=1/2$. While the Son-Dirac CFs are presumably still neutral, the low energy excitations in a quantized Hall state of form $\nu=p/(2p+1)$ are not neutral but have a charge $1/(2p+1)$. Furthermore, the excitations are no longer fermions but obey fractional statistics, only approaching Fermi statistics in the limit $\nu \to 1/2$.\cite{Halperin1984}  

The effects of $F_1$  in the dipolar Fermi liquid are quite similar to the effects of the coupling between the  CFs and the gauge potential $a_\mu$  in the Son-Dirac formulation at $\nu=1/2$. However, the dipolar Fermi liquid picture developed in  the 1990s did not include the Berry phase characteristics featured in Son's formulation. 

\subsection {Trial wave functions and other microscopic theories.}

The focus of the discussion in this chapter has, thus far,  been on effective theories designed to elucidate the low-energy properties of a quantum Hall system at, or very near to, $\nu=1/2$. In general, these theories depend on parameters, such as the effective mass or Fermi velocity, which can only be related to parameters of the original Hamiltonian by other means.  
Methods used to estimate these parameters starting from a microscopic Hamiltonian include fits to results from exact diagonalization of finite sized systems, as well as a variety of approximate calculations.

We have already mentioned the Hamiltonian approach of Ref.~\refcite{rsgm97}   as a way to obtain approximate values for energies. The most
accurate approaches, however,  have made use of ``variational'' trial  wave functions, for which one can calculate numerically  expectation values of the Hamiltonian. Originally introduced by Laughlin to describe FQH states of the form $\nu=1/m$,  trial wave functions were the basis for Jain's composite fermion theory, which gave a natural description for FQH states of the form $p/(2p+1)$, and which were the inspiration for the non-relativistic Fermion-Chern-Simons theory discussed above. (See the chapter by  Jain for more detailed descriptions of these wave functions.) For the case of $\nu=1/2$, the Jain trial wave function takes the form
\be
\label{Jainhf}
\Psi  =  P_{\rm{LLL}}  \Psi_{\rm{FS}}  \prod_{j<k} (z_j - z_k)^2  \prod_l  e^{- |z_l|^2 / 4 l_B^2} ,
\ee
where $P_{\rm{LLL}}$ is a projection operator onto the lowest Landau level,   $\Psi_{\rm{FS}}$ is the wave function for a Fermi sea of non-interacting electrons in zero magnetic field, and $z_j = x_j - i y_j$ is the position of electron $j$ in complex notation. 
Although trial wave functions such as Jain's  are entirely within a single Landau level,  they are not generally the exact ground state of a Hamiltonian with pure two-body interactions, so they need not  obey precisely the requirements of PH symmetry. For example, the two-point correlation function for fluctuations in the electron density obtained from a CF trial wave function at a filling fraction $\nu$ will not be precisely the same as that obtained for filling fraction $1-\nu$ obtained with the same CF prescription.  Nevertheless, numerical calculations  have found that properly chosen trial wave functions give results which are very nearly PH symmetric, just as they give remarkably good agreement with ground state energies obtained by exact diagonalization of  finite systems. \cite{GeraedtsWang2017,balram15}  (See the chapter by Jain for more details.) 

Several recent papers have used model wave functions to look for a Berry phase accumulated by the many-body wave function on a torus when a single CF is moved around a closed path on the surface or interior of the CF Fermi sea at $\nu=1/2$.\cite{WangGeraedts2017,GeraedtsWang2017}.  Using a highly efficient scheme for evaluating properties of a model wave functions based on  a lattice Monte Carlo technique introduced by Haldane, they have been able to study systems with up to 69 electrons. They find that the many body wave function accumulates a Berry phase of $\pi$ when a CF is moved around the Fermi surface, analogous to the Berry phase that would be accumulated from moving a non-interacting Dirac fermion around the Fermi surface in the Son-Dirac scheme.  A $\pi$ Berry phase is also obtained when the CF is moved along a closed path inside the Fermi sea which encloses the origin but does not come too close to it.  The model wave functions employed, however, are based on the non-relativistic CF formulation of  Jain and HLR.    Thus, it would not seem that  this should be interpreted as an argument favoring Son-Dirac over the HLR formulation of $\nu=1/2$, or that there is necessarily a fundamental difference between the two theories. 

It may be worth noting that a recent paper by Pu {\it{et al.}} has studied how the the accumulated Berry phase varies when mixing between Landau levels is included.\cite{PuFremlingJain18}  They find that the Berry phase evolves rapidly from the original value of $\pi$ to a value close to zero, under a relatively small amount of Landau level mixing.

\subsection{Interpretation of Son-Dirac in terms of vortices}
The indirect manner in which the electromagnetic gauge field $A_\mu$  is coupled to the composite fermions in Son's formulation (\ref{dirac}) has led to the interpretation that  the fermions in Son-Dirac should be interpreted to represent  ``vortices" in the electron gas, rather than the electrons themselves, and that the Son-Dirac theory is a low-energy ``dual" description to the original description in terms of electrons. \cite{sonphcfl,WS16,Seiberg2016}     Wang and Senthil \cite{WS16}  explain this point of view as follows.  
If we consider the trial wave function (\ref{Jainhf}) for $\nu=1/2$, then the factor $\prod (z_j - z_k)^2 $ may be interpreted as producing a $4 \pi$ vortex about each original electron.  However, this factor will also 
lead to a suppression of the background particle density in the vicinity of any given electron, such that precisely one electron will be missing from the background at filling factor 1/2.  Therefore, adding in the charge of the chosen electron, the low energy composite fermion will be locally neutral, but it still retains the vorticity. 

The Jain trial wave function for a state of the form $\nu=p/(2p+1)$, with $p>0$,  has a form similar to (\ref{Jainhf}), but with $\Psi_{\rm{FS}}$ replaced by the wave function for an integer quantized Hall state with $p$ filled Landau levels. After projection to the lowest Landau level,  this leads to additional zeroes in the wave function, such that the total number of $4 \pi$ vortices is no longer equal to the number of electrons but is  equal to  one-half the number of flux quanta.  As  the number of fermions in the Son-Dirac formulation is equal to one-half the number of magnetic flux quanta, it is therefore equal to the number of the number of vortices. Thus, it is natural to identify the  underlying fermions in Son's formalism with vortices.

 As originally noted by Read \cite {read94}, the  electric dipole moment of the CFs at $\nu=1/2$ can be understood as arising from a spatial separation of the electron and its screening vortex charge.  Since the direction of the dipole changes as a CF is moved around the Fermi surface, it may not be surprising that a Berry phase will be associated with this process. 
 At other even denominator fractions of the form $\nu=1/2m$, where $m$ is an integer, where the trial wave function has $2m$ zeroes for each electron, one again can argue that the low energy CFs are neutral dipoles, arising from a spatial separation of the electron and the screening vortex charge.  
 Wang and Senthil \cite{WS16} argue that the Berry phase at the Fermi surface should have the value of $\pi/m$ at $\nu=1/2m$.  This coincides with the Son-Dirac picture at $\nu=1/2$, but differs for fractions with $m>1$. Since PH symmetry is no longer present at fractions with $m>1$, it is not surprising that Son's original model is not applicable to those cases. However, a Berry phase different from  $\pi$ can be represented within the Son formalism  if one replaces the massless Dirac fermions in Eq. (\ref{dirac}) by Dirac fermions with an appropriate mass.    Moreover, if, as in (\ref{dirac}),  there is no Chern-Simons term in the action for the gauge field $a_\mu$, then a Berry phase of $\pi/m$ is {\em{necessary}}, in order to satisfy the requirement that states corresponding to integer fillings of the effective Landau levels should correspond to the electron filling factors $\nu=p/ (2mp+1)$, as obtained in the Jain CF construction.
 
 The transformation of a problem involving electrons into a problem involving vortices has a related precedent  in the case of Bose systems. It was proposed long ago, for example,  that the phase transition between the superfluid and normal state of a Bose liquid at finite temperatures in three-dimensions, or at zero temperature in (2+1) dimensions, could be interpreted in terms of a proliferation of vortex lines, and that the vortex lines in turn could be interpreted as world lines for a system of bosons interacting with a non-compact U(1) gauge field. \cite{chandandual,peskin78}
The transformation from particles to vortex-like variables has even earlier precedents in the case of Ising systems, and  is  known as a duality transformation.  In the last few years, there have been many theoretical developments involving duality transitions in Fermi systems, inspired in large measure  by the successes of the Son-Dirac theory.  For example, duality transformations have been used to establish connections between the problem of a half-full Landau level and states that could be realized on the surface of a three-dimensional topological insulator.\cite{cfltislrev,WS16,mrosscdl15,dualdrmaxav,dualdrcwts2015,bulkdualmax}
For further examples, see the discussions of a ``network of dualities" in Refs.~\refcite{Seiberg2016,KarchTong16,SenthilSWX19}.

\section{Other Issues} \label{issues}

\subsection {Issues of anisotropy and geometry}

Thus far, we have focused on systems in which the microscopic Hamiltonian is effectively invariant under rotations, and we have assumed that this isotropy is maintained in the final ground state.  However, recent experiments  have been able to measure distortions of the CF Fermi surface in systems where the electronic Hamiltonian is significantly anisotropic, and have been able to compare anisotropies of the CF Fermi surface with those of the electron Fermi surface, measured or calculated, for zero perpendicular magnetic field. 
The reader is referred to chapter by M. Shayegan  for a discussion of these experiments and references to theoretical efforts to explain them.

A second issue concerns the possibility for spontaneous development of  uniaxial anisotropy, in systems where the microscopic Hamiltonian has little or no anisotropy. Theoretical work carried out in the 1970s, before observations of either the integer or fractional quantized Hall effect, noted that according to the Hartree-Fock approximation, a partially filled Landau level would be unstable to the formation of charge density waves. \cite{FukuyamaPlatzmanAnderson}  Moreover at one-half filling, the likely ground state was predicted to be a single  charge density wave, or stripe pattern, which would break both translational and orientational symmetry, and would lead to a large anisotropy in the electrical properties.  The subsequent discovery of fractional quantized Hall effects showed that the actual ground states, at least in the middle of the lowest Landau level, cannot be properly described by a Hartree-Fock approximation, and the possibility of charge-density waves was therefore largely ignored for more than a decade.  However, in 1996, Koulakov {\it{et al.}}  pointed out that  in higher Landau levels, the correlations responsible for formation of fractional quantized Hall states become much less important, and Hartree-Fock should again become reliable.\cite{koulakovKFS96,foglerFK97} In 1998, experiments by the Eisenstein group at Caltech revealed a strongly anisotropic resistivity in high quality GaAs samples at low temperatures, which was consistent with the formation of  charge-density-wave order \cite{lillyanis99}, for states near half filling of the third and higher Landau levels, {\it{i.e.}}, for states with $\nu$ in the vicinity of 9/2, 11/2, 13/2, {\it{etc}}.   Moreover, it was found that  in the second Landau level,  application of an in-plane magnetic field sufficient to destroy the gapped quantized Hall  states at  $\nu=5/2$ and 7/2,       also led to a strongly anisotropic resistance.   However, continuing experimental and theoretical work suggests that the actual electronic states may be better described, in most cases,  as a {\em nematic}, with uniaxial anisotropy but at most short-range stripe-like translational order. \cite{FradkinK99,FradkinKMN00,OppenHS00,CooperLEJPW01,WexlerD01,RadzihovskyD02,QianM17,NguyenGS18} Theoretical investigations have included questions about the nature of the transition between the  nematic and isotropic states. See also Ref.~\refcite{YouCF16} and references therein.

Considerable theoretical and experimental effort has been devoted in order to identify small microscopic perturbations,  including growth anisotropies or effects of a tilted magnetic field, that would select between one or another of the possible orientations for nematic order. Similarly, there has been considerable effort to study the competition and possible phase transitions between states of nematic order and isotropic states, such as a Fermi liquid or a gapped quantum Hall state near half-filling. This competition is clearly of particular importance in the second Landau level, where different phases have been observed depending on details of the system.   
Experiments have also produced evidence for spontaneously-developed anisotropy in 
gapped fractional quantized Hall states at certain filing fractions.\cite{XiaCEPW10}.   Such anisotropies could be found, for example, in transport properties at finite temperature.  Mechanisms by which anisotropy might develop in a gapped quantum Hall state have also been explored theoretically. See, {\it {e.g.}} Refs.~\refcite{YouCF14,WanY16} and references therein.  The reader is  referred to the chapter by Cs\'athy for further discussion and references on all of these points.

There have also been numerous works in the theoretical literature discussing the response of a quantized Hall state to variations of the metric tensor describing the underlying spatial geometry. The ``shift parameter", which characterizes changes in the number of flux quanta necessary to realize a quantized Hall state with a given number of electrons on a sphere, is a well-known example of such a geometric effect.\cite{wen1995,WenZ92}.  Closely related to the shift parameter is the Hall viscosity, which describes electrical currents that can be induced by a changing metric tensor in a quantized Hall state.\cite{read2009} It has been suggested that metric fluctuations produced by an acoustic wave might be used to excite quadrupolar modes of a gapped quantum state.\cite{yang2016} Geometric aspects of states in the lowest Landau level and their response to anisotropy have been emphasized in various investigations.\cite{IppolitiBH18,haldane2018, LiouHYR19}

It is not entirely clear whether there should be unique definition of the shift parameter for a gapless state such as at $\nu=1/2$, and one must be careful about  the proper definition of the Hall  viscosity in this case.   However,  the Son-Dirac theory has been explicitly formulated in the presence of a general metric tensor, so it could be used to investigate the response to variations in the geometry.\cite{son18}  The HLR approach as originally formulated applied only to the case of a planar geometry, but extensions to a general metric have been recently developed.\cite{ChoYF14,GromovCYAF15}.

\subsection{Thermal effects}

Among the fundamental questions one may ask about a gapless state in strong magnetic field  are  the  behaviors of thermal properties, such as the thermal conductivity and thermoelectric effects, in the limits of low finite temperatures and small impurity densities.   Within either  the HLR or Son-Dirac picture, energy current is carried by CFs, and the longitudinal  thermal conductivity  $\kappa_{xx}$, at $\nu=1/2$,  should be proportional to $C_p  v_F  l_{\rm{cf}}$, with the specific heat $C_p$  proportional to $T$, ignoring possible effective-mass divergences.  By contrast, the electrical conductivity $\sigma_{xx}$ will go to zero proportional to  $1/l_{\rm{cf}}$, for large mean-free-path. It is clear, therefore, that the ratio 
$\kappa_{xx} / T \sigma_{xx}$ will vastly exceed the Wiedemann-Franz value under these conditions.

Thermoelectric effects at $\nu=1/2$ and $\nu=3/2$ were discussed two decades ago, by Cooper et al. \cite{cooperCHR97}  within the HLR picture. 
More recently, Potter et al. \cite{PSV} calculated both the diagonal and off-diagonal components of Seebeck tensor $\hat{S}$, which relates the generated electric field to the heat current in a geometry where no electrical current is allowed,  within the Son-Dirac picture as well as HLR.  While they found identical results in the two formalisms for the leading, diagonal, component of $\hat{S}$, they found a difference between predictions using HLR and Son-Dirac for the Hall component, which is smaller than the diagonal component  by a factor of order $(k_F l_{\rm{cf}})^{-1}$ in a clean system.  

However, there are subtleties here which require further investigation.  On the one hand, the analysis of Ref.~\refcite{PSV} did  not take into account the effects of  fluctuations in the electrostatic potential due to the imperfect screening of impurities, which are important in the calculation of $\sigma_{yx}$, as we have seen.  In addition, there can be a large contribution to the thermal Hall conductance from regions near the sample boundary, where PH symmetry will be strongly broken.  
At this point, it is not completely clear whether there is an actual difference between the two theories in their prediction for the Hall contribution. Recent measurements by Liu et al.\cite{LiuLZWZD18}  of $S_{xy}$ in a GaAs 2DEG at $\nu=1/2$ found values that were smaller than the values predicted by Potter et al. and that fell off  faster with decreasing temperature than the expected linear dependence.

 It should also be noted, that predictions for the ratio between the values of $S_{xx}$  at $\nu=1/2$ and $\nu=3/2$  in  Ref.~\refcite{cooperCHR97} were not found to be in good agreement with experimental data available at the time.
Reference \refcite{LiuLZWZD18} reported measurements of $S_{xx}$ at $\nu=1/2$, whose magnitude was consistent with expectations, but measurements were not reported at $\nu=3/2$, so a comparison cannot be made. 

\subsection {Effects of strong disorder}.

In Subsection \ref{3.1}, we discussed the behavior of the the conductivity tensor at $\nu=1/2$ in the limit of weak, but non-zero, scattering due to disorder.  We argued using either the HLR or Son-Dirac picture, that there should be a value of $\sigma_{xx}$ proportional to the inverse  mean-free-path for CFs,   $l_{\rm{cf}}^{-1}$, which was implicitly assumed to be small, proportional to the impurity concentration for weak disorder.  However, this analysis was based on the use of a Boltzmann equation, treating the CFs as non-interacting fermions in an a random effective magnetic field (and correlated electrostatic potential),  which we do not expect to be strictly correct in the limit of zero temperature. As pointed out in Ref.~\refcite{hlr}, a diagrammatic analysis of the effects of fluctuations suggests that their should be  corrections to $\sigma_{xx}$ proportional to $\ln (T \,  l_{\rm{cf}} /v_F)$, which, depending on their sign, should drive  $\sigma_{xx}$ towards zero or to a value of the order of $1/4 \pi$ in the limit of $T \to 0$. In practice, these corrections may be negligible for experiments on the highest-quality samples, but they are  of interest from a theoretical point of view. In any case, in samples of lower quality, one can certainly encounter values of $\sigma_{xx}$  that are not small compared to $1/4 \pi$, and we would like to understand what can happen in that case. 

\begin{figure}[hb]
	\centerline{\includegraphics[width=6.7cm]{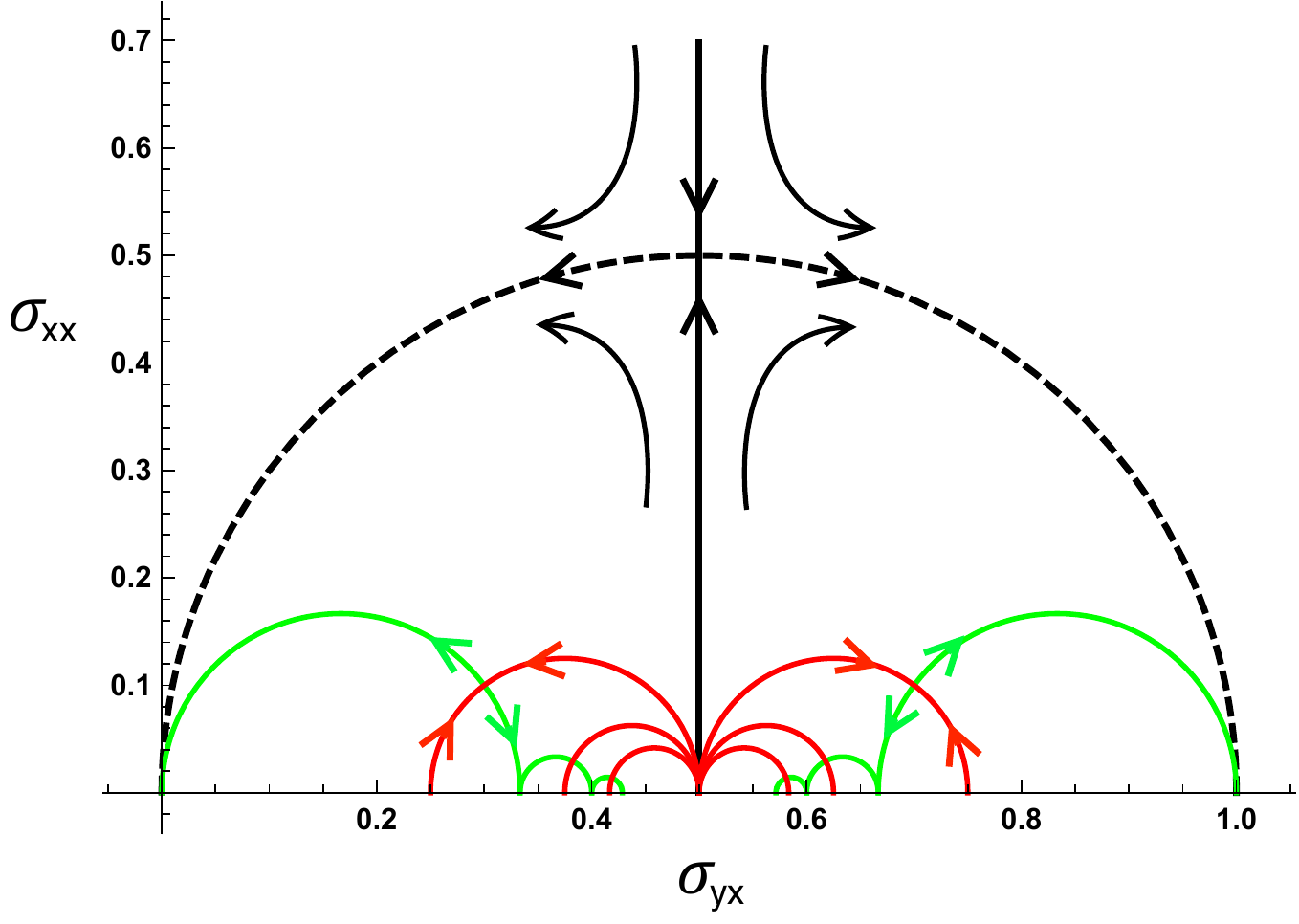}}
	\caption{Proposed renormalization-group flow in the plane of $\sigma_{yx}$ and $\sigma_{xx}$, in units of $e^2/h$. The vertical solid black line separates points which flow, at large length scales, into the insulating fixed point, with $\sigma_{xx} = \sigma_{yx} = 0$ and the integer quantized Hall state with $\sigma_{yx}=e^2/h$.
Points on the vertical solid black line flow into the unstable fixed point at   $\sigma_{xx} = \sigma_{yx} = e^2/2h$. 
The dashed black curve shows the flow trajectories from the unstable fixed point to the stable fixed points at $\sigma_{xx} = 0$.  Red curves separate a region which flows to  a fractional quantized fixed point from a region which flows to another fraction or to integer quantized fixed points. Green curves show  trajectories flowing from the unstable fixed point on a red curve into a quantized fixed point at $\sigma_{xx} = 0$. Only six fractional state are shown, with $\nu = 1/3, 2/5, 3/7, 4/7, 2/5$ and 2/3.  }
	\label{RGflow}
\end{figure}

In 1983, Khmel'nitski\u{i}  \cite {khmelnitskii83}, proposed a theory of phase transitions between integer quantized Hall states in the presence of disorder, based on  a
 renormalization group (RG) involving two parameters, $\sigma_{xx}$  and $\sigma_{yx}$.  He argued that the RG  should have a periodic dependence on  the variable $\sigma_{yx}$, with stable fixed points corresponding to the quantized Hall states, with  $\sigma_{xx}=0$ and $\sigma_{yx}$ equal to an arbitrary integer multiple of $1/2\pi$, and that there should be a set of unstable fixed-points with 
\be
\label{fixedpoints}
\sigma_{yx} = \frac {n+1/2} {4 \pi} , \,\,\,\,\,  \sigma_{xx} =  \sigma_{c},
\ee 
where $n$ is an arbitrary integer, and the critical conductance $\sigma_c$  should be independent of $n$ and of order $1/4 \pi$.  In addition, there should be an unstable  fixed point at $\sigma_{xx}= \infty$.  According to this analysis, if one starts with any value of $\sigma_{yx}$ that is not of the form $(n+1/2)/4 \pi$,
one  should flow at large length scales to the nearest integer quantized Hall state, with $\sigma_{xx}=0$. On the other hand, if one starts with a value of $\sigma_{yx}$  that is precisely equal to $(n+1/2) / 4 \pi$, the value of $\sigma_{yx}$ should remain constant, and one should flow to the associated fixed point with $\sigma_{xx} = \sigma_c$ and the initial value of $\sigma_{yx}$.   
The qualitative flow pattern implied by this analysis is illustrated by the black lines and arrows in Fig.~\ref{RGflow}. 
[Note: The RG flows are invariant under the interchange of $\sigma_{xy}$ and $\sigma_{yx} = - \sigma_{xy}$. We use $\sigma_{yx}$ in our discussions, because it is positive for electrons with $B>0$. ]

Khmel'nitski\u{i}'s picture, and subsequent work by Pruisken \cite{pruisken88},
did  not take into account 
 the existence of fractional quantized Hall states, which one should find  in interacting electron systems of sufficiently high quality. Kivelson, Lee, and Zhang\cite{kivelsonLZ92}   proposed  a generalization of the phase diagram, motivated by Jain's CF picture, which included fractional quantized Hall states at arbitrary odd-denominator fractions.   RG flow lines associated with some of these states and the allowed transitions between them are indicated by colored curves in Fig.~\ref{RGflow}.  The description of transitions between the various integer quantized Hall states is not affected,  as the unstable fixed points  (\ref{fixedpoints}) are not in the region of the new flow lines.  Regions  of the phase diagram that flow to fractional quantized Hall states can only be accessed if the disorder is not too strong, and if the starting value of $\sigma_{yx}$ is not too close to $(n+1/2)/2 \pi$.
 The fixed points of the RG flow represent possible values of the conductivity tensor in the limit of an  infinite system at $T=0$.   At non-zero temperatures, or in a finite system, the RG flow can stop before one reaches a fixed point, so other values of the conductivity are possible.  
Because the RG flow rates are slow in some regions of the phase diagram, there will typically be ranges of the magnetic field, in any sample, where the  values of $\sigma_{yx}$  and   $\sigma_{xx}$  vary continuously with filling factor, and are almost independent of temperature, to the lowest attainable temperatures.    We  may describe such a region as an intermediate  {\em{metallic phase}}.

According to the flow diagram illustrated in  Fig. \ref{RGflow}, direct transitions are not allowed between arbitrary  pairs of quantized Hall states   In particular, at $T=0$,  
it is not possible to have a direct transition between a fractional quantized Hall plateau  with $\nu <1/2$ and one with $\nu >1/2$, as one varies the electron filling factor.   One would have to first pass through 
the insulating state at $\nu=0$ and the integer state at $\nu=1$, which we would describe as {\em {re-entrant}} integer states.    Of course, for a finite system or at finite temperatures, the  fractional quantized Hall plateaus could be separated by an  intermediate metallic state rather than by re-entrant quantized  Hall states.  [{\it{Cf.}} Ref~\refcite{hlr}.]

Beyond the location of fixed points and phase boundaries, an RG calculation should also make predictions for the {\em{rates}} at which physical properties evolve as a function of increasing length scale or decreasing temperature. If one starts with $\sigma_{xx}=\sigma_c$ and $\sigma_{yx} $ very close to $1/4\pi$, the renormalization group analysis picture  says that the deviation $\Delta \sigma_{yx} \equiv  \sigma_{yx}- (2n+1)/4 \pi $ should grow with the measuring  length scale $L$ as $L^{1/\tilde{\nu}}$,  where $\tilde{\nu}$ is  a critical exponent determined by form of the $\beta$- function near the critical point, while $\sigma_{xx}$ should remain roughly constant as long as  $|\Delta \sigma_{yx}|$ remains small. (We denote the exponent as $\tilde{\nu}$ to distinguish it from the filling factor $\nu$.) 
For an infinite system at non-zero temperature, the
 conductivity tensor  should be  determined by stopping the renormalization group flow at a length scale $L_T$ determined by the temperature, following a power law which may be written in the form $L_T  \propto T^{-z_T}$.   
 The exponent $z_T$ is supposed to arise from the temperature dependence of the mean-free-path for inelastic scattering.\cite{pruisken88} If  the filling factor is varied through the value 1/2, while the temperature is fixed,  $\sigma_{yx}$ should undergo its transition from a value close to zero to  a value close to $1/ 2 \pi$ over an interval  $\Delta \nu^*$ of filling fraction proportional to   $T^\kappa$, where, following the RG, $\kappa = (z_T \tilde{\nu})^{-1}$. The precise values of these exponents have been the objects of continuing theoretical and experimental studies.

To obtain a theoretical estimate of the exponent $\tilde{\nu}$, the RG picture   can be compared with results for the properties of a non-interacting electron in the lowest Landau level in the presence of a random potential.  In this case it has been found that  all  eigenstates are localized, except at  a critical energy $E_c$, which will sit in the middle of the Landau level in the case of disorder with PH symmetry. The localization length $\xi$ is found to diverge as the energy $E$ approaches $E_C$, and it is expected that the divergence should have the form of a power law, 
$\xi \propto |E-E_c|^{- \bar{\nu}} $. It has been argued that short-range electron electron interactions will appear as an irrelevant perturbation at the RG fixed point describing the critical point of the non-interacting system, and consequently, for short-rage interactions,  the exponent $\tilde{\nu}$ should be identical to $\bar{\nu}$.\cite{BurmistrovBEGM11} On the other hand, Coulomb interactions, which fall off as $1/r$, are found to be relevant at the non-interacting fixed point, so the value of $\tilde{\nu}$ may be different from the non-interacting $\bar{\nu}$ in that case. The theoretical discussions also assume that there are no long-range correlations in the disorder potential or macroscopic inhomogeneities in the system.   Long-range-correlated disorder can convert the problem to a classical percolation problem, which certainly has a different values of the critical exponents.

Many of the numerical studies have employed a simplified model, introduced by Chalker and Coddington (CC) in 1988,\cite{ChalkerCoddington} in which regions of with quantized Hall conductances $\nu=1$ and $\nu=0$ form a regular checkerboard pattern. Electrons flow along the edges  between the squares in a definite direction, such that each vertex has two incoming and two outgoing bonds.  Scattering at a node is characterized by an S-matrix, in which there is a probability $p$ to scatter to the right and $1-p$ to scatter to the left.  In the CC model, the probability $p$ is assumed to be the same at every vertex, while the phase accumulated by an electron traveling along a link is   assumed to be random. The transition point between states with macroscopic quantized Hall conductance occurs at $p=1/2$.  Although early simulations of the CC model gave results for $\bar{\nu}$ of order 2.35, more recent studies of larger systems give results closer to 2.6. (See, {\it {e.g.}}, Refs.~\refcite{ObuseGE12,NudingKS15,FulgaHAB11}.)  More recently, Zirnbauer has argued that the correct asymptotic value for the CC model  should be $\bar{\nu} = \infty$, and that apparent values in the range of 2.3 to 2.6 reflect  slow flow by a marginal variable towards the asymptotic fixed point.\cite{zirnbauer2019}. 
Moreover, Zirnbauer calls  into question the entire validity of a two parameter RG as a description of this problem. 

Regardless of the interpretation, it appears that there is a discrepancy between the values of $\bar{\nu}$ obtained from numerical studies fo the CC model and the value of $\tilde{\nu }\approx 2.38\pm 0.06 $ extracted from experiments.\cite{LiVXPTPW09}  One possibility is that the difference is due to effects of electron-electron interactions that are missing from the CC model. It is also possible that  long-range correlations in the impurity potential are affecting the experimental results. However,  Gruzberg {\it {et al.}} have suggested that the problem arises from the the regular geometric arrangement of scattering vertices in the CC model. They have studied numerically a set of models in which some fraction of the vertices have scattering probabilities  $p=0$ or $p=1$, effectively removing them fro the network, so that the remaining vertices are connected in an irregular way.\cite{GruzbergKNS17,KlumperNS19} According to their analysis,  the exponent obtained for  their model is   $\bar{\nu} \approx 2.37$, in good agreement with experiment.

 Beyond the value of $\bar{\nu}$. many theoretical investigations have been concerned with the multifractal structure of the electron wave function at the critical point for the non-interacting problem. (See, {\it {e.g.}}, Refs. \refcite{BurmistrovBEGM11} and \refcite{zirnbauer2019}). This structure is probably  not accessible in experiments but theory  can be compared with numerical simulations.

Since  the value of $\tilde{\nu}$ for system with short-range interactions is identical to the value of $\bar{\nu}$ for a non-interacting electron system, the size-dependence of the conductivity tensor for a finite system at $T=0$ should be similar for the two systems. 
However, the behavior   at finite temperature will generally be different in the two situations.  
For a non-interacting system, the conductivity tensor at non-zero temperature $T$ will be equal to the conductivity of the same system at $T=0$,  but averaged over  Fermi level,  weighted by  the derivative of the Fermi function at temperature $T$. Since states of the noninteracting system are localized at all energies except $E_c$, they do not contribute to $\sigma_{xx}$, and  the resulting longitudinal conductivity for $T \neq 0$ will be zero, regardless of the chemical potential $E_F$.  If the value of $E_F$ is allowed to vary,  
 $\sigma_{yx}$ will be vary smoothly between a value $n/ 2 \pi$, for $E_F  \ll E_c$,    and  the value $(n+1)/2 \pi$, for 
 $E_F  \gg E_c$, over a range of filling factors $\Delta \nu^*$ that will be of order $T/W$, where $W$ is the width of the Landau level due to the disorder.  This may be contrasted with the RG prediction for interacting systems that $\sigma_{yx}$ and  $\sigma_{xx}$ should both be finite and vary smoothly over a range of the form 
  $\Delta \nu ^* \propto T^\kappa$, with $\kappa=(z_T \tilde{\nu})^{-1}$.  According to Ref. \refcite{BurmistrovBEGM11}, the value of $z_T$ for short-range interactions should be $z_T = 1.23$, suggesting a value of $\kappa$ in the range 0.314 to 0.346.  However, experiments obtain a value $\kappa = 0.42 \pm 0.01$,\cite{shaharTSSS97,BurmistrovBEGM11}     which would imply $z_T \approx 1.0 \pm 0.1$.  It is not known theoretically what the value of $z_T$  should be in the case of Coulomb interactions, and it appears beyond current capabilities to obtain a reliable value from computer simulations.

In the  paper by Kivelson {\it{et al.}}\cite{kivelsonLZ92} mentioned earlier, it was suggested that at $T \neq 0$, in the vicinity of the transition point between insulator and integer quantized Hall conductor, the Hall resistivity $\rho_{yx}$ should remain constant, at a value equal to $2 \pi$, throughout the region where $\rho_{xx}$ changes from a value $\gg 1/ 2\pi$ to a value $ \ll 1/ 2 \pi$.  This constancy of $\rho_{yx}$  is equivalent to the statement that  the complex conductivity should fall on a semicircle, 
\be
\label{semicircle}
\sigma_{xx}^2 + (\Delta \sigma_{yx})^2 = (1/4 \pi)^2 .
\ee
Invoking PH symmetry, we expect that at a fixed temperature,  $\sigma_{xx} (\nu)$ should be symmetric about $\nu =1/2$, while $\Delta \sigma_{yx}$ should be antisymmetric. This implies  that  at a fixed non-zero temperature $T$, the resistivity
 near the transition, at a distance $\Delta \nu$ from the critical value,  should obey the {\it {duality relation}}   
\be
\label{rhodual}
\rho_{xx} (\Delta \nu) = \frac{(2 \pi )^2}{\rho_{xx} (- \Delta \nu)} 
\ee
These equations  also imply $\sigma_c = 1/4 \pi$.
The predictions  (\ref{semicircle}) and(\ref{rhodual})
are well-supported by  experiments. \cite {shaharTSSS97}
(By contrast, the analysis of Zirnbauer\cite{zirnbauer2019} predicts $2 \pi \sigma_c  = 0.6366...$  .)

Recently, Kumar, Kim, and Raghu \cite{Kumar19} have analyzed the transition between the $\nu=0$ insulating state and the $\nu=1$ quantized Hall state by comparing the formulation in terms of electrons  with a formulation based on CFs, using both the HLR and Son-Dirac formulations.  They find a duality between the two descriptions, which leads to a self-duality at the critical point and, hence, to the prediction  $\sigma_c = 1/4 \pi$. 
The analysis of Ref.~\refcite{Kumar19} assumes one can ignore residual interactions between the CFs, though it does not assume validity of a Boltzmann equation. However, it is substantiated by comparisons with an analysis using a generalized Chalker-Coddington-type network model for the CFs, as well as one based on a non-linear sigma model.

In earlier work,  Dolan \cite{DOLAN1999}  proposed an analytic form for the flow diagram, based on several assumptions, which implied Eqs.~(\ref{semicircle}) and (\ref{rhodual}), as well as 
 $\sigma_c =1/4 \pi$.  One assumption was that the diagram should be consistent   
with an assumed symmetry  expressed in terms of the action of the modular group on the complex conductivity, $\sigma_{yx}+ i\sigma_{xx}$, in the 
upper-half  plane. (This assumption may be understood as a natural extension of the transformations by flux attachment employed by Jain.)   A second key assumption was  that the $\beta$-function for the complex conductivity, which describes the flow of parameters under a change of length scale, should be a meromorphic function of the complex conductivity, (after multiplication by a real function of $\sigma_{xx}$ and $\sigma_{yx}$). 
This assumption was motivated by an analogy with apparently unrelated work on Yang-Mills gauge theory in four-dimensions, but was not derived from a microscopic theory of quantum Hall systems.  The assumption of modular invariance implies that if the flow trajectories  from the unstable  point $\sigma_{xx} = \sigma_{yx} = 1/4 \pi$ to the stable fixed points  with $\sigma_{xx}=0$ and $\sigma_{yx} = 0$  or $1/2\pi$ lies on the semicircle (\ref{semicircle}), and the boundary between regions which flow to the two stable fixed points is given by the  line  $\sigma_{xx} = 1/ 4 \pi$, then the trajectories and phase boundaries  related to fractional quantized Hall states, represented by the colored curves in Fig.~\ref {RGflow} will all be semicircles.

It should be emphasized  that a  renormalization group analysis cannot, by itself, determine the detailed phase diagram of a real physical system. A $\beta$-function  based on two parameters will apply only at length scales large compared to any microscopic length, so other variables must be considered at short distances.  Some simplification occurs if one assumes that both the disorder potential and electron-electron interaction are much smaller than the cyclotron energy, so that Landau level mixing can be neglected even in the presence of disorder, and one assumes that the disorder is PH symmetric. Then if the Landau level is precisely half-full, the value of $\sigma_{yx}$ is fixed at $1/4 \pi$ at all length scales and temperatures. Then the transition between integer quantized Hall plateaus at $T=0$ will occur precisely when the filling factor passes through 1/2. When the disorder potential is strong enough  so that Landau level mixing needs to be taken into account, the transition will generally be shifted to larger values of the filling factor.  For very large disorder, the transition could be shifted to very high filling factors, so that the system remains localized and  is insulating at $T=0$ for any achievable electron density and magnetic field strength.

\section {Open questions about $\nu \approx 1/2$}. \label{open}

Discrepancies between the predictions of HLR and Son-Dirac at the RPA level, such as those discussed  above in Subsection \ref{disagree}, raise the question whether the two theories are fundamentally different, or whether they will become equivalent when corrections beyond RPA are taken into account.  
If one goes beyond the RPA level, both the HLR theory and the Son-Dirac theory contain an infinite number of parameters, such as vertex corrections, to account for fluctuations at non-zero wave vectors and frequencies. (Vertex corrections mght include corrections to the interaction between CFs and the gauge fields at finite wavevector, as well as  corrections to the interactions between CFs and perhaps wavevector dependent corrections to the Lagrangian for the gauge fields themselves. The parameters describing these corrections  will generally depend on details of the microscopic Hamiltonian, but we do not know how to calculate their values for any particular model. Based on our investigations, it  appears possible to choose values for the vertex corrections in the HLR theory that will remove the discrepancies cited in Refs. \refcite{NguyenGolkar17}  and \refcite{LevinSon}, but it is not clear whether these corrections can be chosen so that they are compatible with Gallilean invariance and other microscopic properties of the starting Hamiltonian (\ref{start}) or (\ref{Lhlr}). This remains an open problem for future investigation.

If it turns out that there is no way to reconcile the HLR theory with PH symmetry, the question will be to understand where HLR goes wrong, when it seems to give the correct answer to so many problems.  The basic assumption that  correct low-energy theory can be obtained by an adiabatic continuation from the mean-field HLR ground state could turn out to be false, but one might then expect many more discrepancies than we have found.

While the Son-Dirac theory, together with regular vertex corrections allowed by symmetry, together with proper account of the infrared divergences discussed  in Subsection \ref{divergences},  appears to be a correct theory of the behavior near $\nu=1/2$ for a system restricted to the lowest Landau level,  it is currently not clear how one can derive the theory from a microscopic Hamiltonian. This also is an unsolved problem.

At the {\em physical} level, we may distinguish between two types of properties.  On the one hand, there are properties which should display PH symmetry asymptotically close to $\nu=1/2$, regardless of whether the microscopic Hamiltonian has PH symmetry or not.  Examples that we have discussed include the values of $\nu$ associated with quantized fractional Hall states, and the energy gaps at these states, at least to lowest order in $|\Delta B|$.  Based on the agreement between HLR and Son-Dirac at the RPA level, it seems most likely that this will also be true for properties such as the peak positions of the Weiss oscillations at order $|\Delta B|^2$, or the value of the Hall conductance at $\nu=1/2$, to order $l_{\rm{cf}}^{-2}$.  Calculation of these quantities within HLR did not depend on the value of the Landau parameter $F_1$, and therefore did not assume restriction to a single Landau level. 

Clearly, there must be other properties which do depend on the presence or absence of PH symmetry in the underlying microscopic Hamiltonian.  In fact, one would expect that virtually any quantitative property would manifest an absence of PH symmetry at sufficiently high order in $\Delta B$, when there is mixing between Landau levels.  We may speculate that the quantities where one finds a discrepancy between HLR and Son-Dirac are of this type, {\em i.e.}, these quantities depend on microscopic details and should agree with the Son-Dirac prediction only in the case where one can neglect Landau level mixing. If the discrepancy between HLR and Son-Dirac can be repaired by inclusion of appropriate vertex corrections, then these corrections should depend on the microscopic details. 

On a broader level, there remain general,  perhaps philosophical, questions about how much accuracy one should expect from a low-energy theory of a system  with gauge fields and a Fermi surface.

If it turns out, in the end,  that the HLR approach is fundamentally incompatible with PH symmetry, then several possibilities remain. One is that HLR is simply wrong, for reasons that we do not yet understand. A second possibility, which has been suggested, \cite{Barkeshli15}  is that for a system restricted to the lowest Landau level,  HLR describes correctly the states with $\nu < 1/2$, but not  states with $\nu>1/2$.  For the latter states, one would need to first  make a PH transformation, apply HLR as a description of the holes, and then transform back to electrons. Exactly at $\nu=1/2$, this would imply the existence of two inequivalent ground states, distinguishable by an appropriate local measurement. In view of the emergent PH symmetry obtained in HLR  at the RPA level for a wide variety of properties, as well as the nearly perfect emergent PH symmetry observed in calculations based on composite fermion trial wave functions,
this seems like a rather unlikely possibility, and there is no  evidence for such a situation in any numerical calculations that we are aware of.


\section{Gapped Quantized Hall States at Even-Denominator Fractions} \label {gapped}

Although there is no indication of an energy gap or quantized Hall plateau at filling factors 1/2 or 3/2 in narrow-well GaAs samples, quantized Hall plateaus at a half-filled Landau level have been observed in other situations.  Best known are the quantized Hall states at $\nu=5/2$ and $\nu=7/2$ in GaAs structures, which are discussed in several chapters of this book. (See the chapters by Heiblum and Feldman, by  Csathy, and by Zalatel.)   An assortment of even-denominator fractional quantized Hall states have also been seen in graphene structures, as discussed in the chapter by Kim, Dean, Li, and Young,  and in ZnO structures, as discussed in the chapter by Falson and Smet.  The precise natures of these various states are still under debate. 

The quantized Hall state at $\nu=5/2$ in GaAs  was first  observed by Willett et al in 1987.\cite {willetteven87}  In 1991, Moore and Read proposed a microscopic wave  function, involving a Pfaffian  of  the variables $(z_i - z_j)^{-1}$,
where $z_i$ is the position of electron $i$ in complex notation,
which provided a possible explanation for the 5/2 state. \cite {MooreMR91} From the beginning, it was realized that the Pfaffian state could be understood in a composite fermion picture, if the Fermi surface became gapped by the formation of Cooper pairs, analogous to p-wave superconductivity.  The connection to p-wave superconductivity, and its topological implications, was further elucidated, most notably, in the 2001 work by Read and Green. \cite{Read&Green}. In 2007, it was realized that the PH conjugate of the Pfaffian state, denoted the anti-Pfaffian, which necessarily has the same energy as the Pfaffian in a model restricted to a single Landau level with purely two-body forces,  is actually a topologically distinct state.\cite{levinapf,ssletalapf}.  Therefore, if either of these states is the correct ground state in a model with PH symmetry at the microscopic level, it means that this symmetry must be spontaneously broken for the ground state at half filling. 

 Numerical calculations based on exact diagonalizations of  finite systems  and using the density-matrix renormalization group  have strongly supported the idea that exact ground state at half-filling with  interaction parameters appropriate to the second Landau level should have good overlap with  either the Pfaffian or  anti-Pfaffian state and should fall into the corresponding topological class.
 \cite{MorfED, Feiguinetal, WangShengHaldane, StorniMorfDasSarma, RezayiSimon, PapicHaldaneRezayi, Zaleteletal, RezayiPRL}   Calculations which take into account PH symmetry breaking effects due to Landau level mixing have tended to favor the anti-Pfaffian over the Pfaffian.\cite{RezayiSimon, Zaleteletal, RezayiPRL,SimonIZR19} However, as discussed in the chapter by Heiblum and Feldman,  recent experimental results have failed to agree with predictions for either the Pfaffian or anti-Pfaffian!  Rather, they seem consistent with another proposed state, the PH-Pfaffian.\cite{sonphcfl}
 
 The various states mentioned above can all be described in terms of BCS pairing of CFs, in either the HLR or Son-Dirac language. Within HLR, the Pfaffian, PH-Pfaffian, and anti-Pfaffian are described, respectively, as arising from pairing in channels with angular momentum  $l$ equal to 1, -1, and -3.  Within Son-Dirac, the Pfaffian, PH-Pfaffian, and anti-Pfaffian are described, respectively, as arising from pairing in channels with angular momentum  $l$ equal to 2, 0, and -2.  The PH symmetry between Pfaffian and anti-Pfaffian states is manifest in the Son-Dirac description, but it is not inconsistent with HLR. What is required, there,  is that the interaction between CFs should have the same strength in the $l=1$ and $l=-3$ channels.
 
 At least within the Son-Dirac formalism, it might seem that the PH-Pfaffian should be  a more natural candidate to explain the 5/2 state than either the Pfaffian or anti-Pfaffian.  However,  in numerical calculations, no trial wave function in the same topological class as the PH-Pfaffian has been found to have reasonable overlap with the exact ground state of a finite system, in the parameter regime where one finds a gapped state at half-filling. 
 Moreover, numerical studies in the spherical geometry do not find evidence for an incompressible state at the shift corresponding to the PH Pfaffian.
 Indeed, it has been proposed that a gapped ground state in this class is impossible for a PH symmetric model restricted to a single Landau level.\cite{Milovanovic} 
 
  As discussed in the chapter by Heiblum and Feldman, there have been suggestions that a state in the class of the PH-Pfaffian could be stabilized by effects of disorder and/or strong Landau level mixing, which have been generally ignored in numerical calculations\cite{feldman}.  Several papers have explored the possibility that such a state could arise in an inhomogeneous system with a mixture of Pfaffian and anti-Pfaffian region, but it appears unlikely that such an inhomogeneous situation can occur in practice.  
\cite{SimonIZR19} 

Recently, experiments on single-layer graphene have demonstrated the existence of gapped  quantized Hall states in the $N=3$ Landau level,  at filling factors 21/2, 23/2 25/2, and 27/2. \cite{KimBTWJSmet19}   Based on numerical calculations, it was argued that the states could be best explained using a ``221 parton construction", originally formulated by Jain, in 1989. \cite{jain89b} In terms of its topological properties, this state is identical to a BCS state of CFs with pairing in the channel $ l = 3$, in the language of HLR, or $l=4$ in the Son-Dirac formulation.
Alternatively, the state might be explained by the PH conjugate of the 221 parton state, which would be described by  pairing in  the  $l= -5$ channel in HLR or $l=-4$ in Son-Dirac. 

\subsection{Non-Abelian Statistics}

As shown originally by Moore and Read, a  striking property of the Pfaffian state is the that the elementary charged excitations from the state should exhibit non-abelian statistics. This would also occur in the anti-Pfaffian, PH-Pfaffian, and the 221 parton states discussed above, and in various other states  that could conceivably be found in  a half-filled Landau level, but not in all conceivable states.  
States with non-abelian statistics of various types have also been  proposed to occur in states at other filling fractions. 

In this chapter, we shall only review a few key points related to the experimental search for non-Abelian statistics at $\nu=5/2$.  
A more complete discussion of non-abeilan statistics in  various gapped fractional quantized Hall states may be found in several other chapters of this book, including the ones by Simon, by Stern, and by Heiblum and Feldman.

In all of the states that have been proposed for $\nu=5/2$, the elementary charged excitations are supposed to have charge $\pm 1/4$.  For the states supporting non-abelian statistics, such as the  Moore-Read,   if there are an even number $n$ such quasiparticles that are each localized in space, say by pinning to charged impurities that  are sufficiently well separated from each other, then the ground state of the system will be a Hilbert space of dimension $2^{n/2}$.  [The Hilbert space dimension will be  $2^{(n+1)/2}$ if $n$ is odd.]
In a finite system, the energy levels in this Hilbert space will not be truly degenerate but will have energy splittings that fall off exponentially with the separation between quasiparticles and with their separations from the boundary. If, by varying the Hamiltonian, one move the quasiparticles around in such a way as to  interchange their positions  and/or wind them around each other, at a rate which is adiabatically slow on the scale of the energy gap to all other excitations, and if the set of quasiparticle positions at the end is the same  as it was at the beginning, then the system will remain within the manifold of ground states and new ground state will be related to the old by a unitary transformation. 
If, at the same time, the interchange is fast compared to the exponentially small energy splittings between the ground states, the unitary transformation will be independent of all details other than the space-time topology of braiding procedure. In general, the unitary transformation will depend on the order in which quasiparticles are interchanged, so that  the set of transformations forms a non-abelian representation of the braid group.

The non-abelian properties of this system can be described by saying  that there is a localized zero-energy Majorana mode attached to each quasiparticle in the bulk, as well as a set of propagating Majorana modes at the boundary. In the case where the number $n$ of enclosed charge $\pm 1/4$  quasiparticles is odd, one of the boundary Majorana modes will be precisely at zero energy; if $n$ is even, then the lowest boundary mode will have a finite energy, inversely proportional to the length of the boundary.

An alternative to the physical braiding of localized quasiparticles is to carry out a Fabry-Perot-type interferometry experiment, as described in the chapter. of Heiblum and Feldman. It is predicted that due to the non-abelian statistics, if the area of the interferometer is held fixed  while the magnetic field is varied, and no quasiparticles are allowed to enter or leave the interferometer region, then the interference pattern would have  different periodicities in the cases where the enclosed quasiparticle number is even or odd.  In practice, one would expect the quasiparticle number itself to vary as the magnetic field is changed, so one is led to an interference pattern that alternates between two periodicities.  In a series of papers dating back to 2007, Willett and coworkers have reported observations of such an even-odd alternation in carefully prepared GaAs samples, at filling factors 5/2 and 7/2. \cite{willett2007}   Most recently, they have reported extensive measurements on eleven different samples, with careful analyses of the oscillatory dependences of the observed interferometer resistance on magnetic field and  on voltages applied to gates designed to change the area of the interferometer. \cite{willett2019} The results appear to give strong support to the existence of non-abelian quasiparticles of the expected type.

Nevertheless, there are aspects of these experiments that are not well understood.  The interferometer areas needed to fit the observations were of order 0.25 $\mu$m$^2$, while the lithographic structures were squares ranging from 2.5 to 5.7 $\mu$m on a side. Also, since interferometer area must presumably connect the openings in the defining gates of the two sides of the interferometer region, the width in the perpendicular direction must be less than 0.1 $\mu$m.   It is not clear what physical factors could give rise to a stable interferometer area of such a small size in these samples. There may also be questions about the extent to which it is appropriate to talk about the existence of a $\nu = 5/2$ quantized Hall state in a region of these dimensions.  

On a more general level, it may be worth emphasizing that while the energy splitting between different ground states for fixed locations of the quasiparticles is supposed to fall off exponentially with the separation between quasiparticles, the expected decay length $\xi$ is not particularly small. One would $\xi$  to be at least a few times larger than the magnetic length, which is approximately 1 nm at a field of 5 T. 
 
 An experiment that wanted to perform topologically controlled braiding by interchanging the positions of two localized quasiparticles would have to  keep the quasiparticles sufficiently far away from each other and from all other quasiparticles  that the energy splitting is small compared to the inverse of the braiding time.  If the pre-exponential factor in the energy splitting is of the order of the energy gap in the 5/2 state (say 200 mK) and if the obtainable  braiding time were of order 10 nsec, the minimum separation would have to be greater than $5\xi$. 
 However, the requirements for a Fabry-Perot interference experiment may be less severe, and interactions between quasiparticles may actually be helpful in stabilizing the interference pattern.\cite{willett2019,RosenowHSS08}


\section{Two-Component Systems: Spin-Polarized Bilayers } \label {bilayers}


We consider here a system of two parallel layers separated by an insulator, such that tunneling between layers may be neglected. We assume that the electron spins are fully aligned, so we can neglect the spin degree of freedom.
Let us  introduce a pseudospin variable $\tau = \pm 1$ to distinguish the layers, and let us  introduce a set of three Pauli matrices $\vec{\tau}$ that operate on the pseudospin index. Since  the electron number in  each layer is separately conserved in our system, the operator $\tau_z$ commutes with the Hamiltonian, and the global U(1) symmetry generated by this operator is a symmetry of the problem.   Nevertheless, in the presence of Coulomb interactions, this rotational symmetry may be spontaneously broken in the ground state, as we shall see. 

In general, the  electron-electron interactions will differ for two electrons in the same or opposite layers.  Typically, the Coulomb repulsion will become independent of the layer index for in-plane electron separations large compared to the distance between layers, while at small in-plane separations the repulsive potential between electrons in different layers will be weaker than for electrons in the same layer.

In the limit of zero separation between the planes, where the interactions are independent of the layer indices, the Hamiltonian becomes identical to that of a single spin-degenerate layer, where the Hamiltonian has an SU(2) spin symmetry.   In that case, if mixing between Landau levels can be neglected, the ground state at total filling $\nu_{\rm{tot}}=1$ can be found exactly.  The solution
is given precisely by the Hartree-Fock approximation and may be described as a ferromagnetic Slater determinant state,  with all spins pointing in the same direction of space.  As there is precisely one electron per flux quantum, the state is 
a filled Landau level for electrons in the selected spin state, so it is essentially an integer quantized Hall state. The favored spin direction is arbitrary in the absence of a Zeeman field, but even a weak Zeeman field will be sufficient to fix the direction of polarization. It should be noted, however,  that though the ground state is completely spin aligned, low-energy Goldstone modes involving spin fluctuations can be important at finite temperatures.\cite{BychkovIE1981,KallinH1984} Moreover, coupled spin-charge excitations  can have dramatic effects on the behavior of a spin-degenerate single-layer system slightly away from $\nu=1$.\cite{SondhiKKR1993,FertigBCM1994,BarrettDPWT1995} Related phenomena can also occur in systems where carriers sit in  two degenerate values  within the two-dimensional Brillouin zone.\cite{RasoltPM1985,RasoltHV1986} 

For the bilayer case we are interested in, Hartree-Fock is no longer exact, but it is still
 expected to be a good approximation when the separation between the layers is sufficiently small compared to the 
magnetic length $l_B$, which is also roughly the separation between neighboring electrons in a single layer.   The Hartree-Fock ground state will be a filled Landau level, with all electrons having the same pseudospin polarization, described by a vector $\sex{\vec{\tau}}$ of unit magnitude.  If the electron concentration is separately specified in the two layers, then the value of $\tau_z$ will be fixed. If there are electrons in both layers, the magnitude $|\sex{\tau_z}|$ will be less than 1. Then, the component of  $\sex{\vec{\tau}}$ perpendicular to the axis will be non-zero, and can point in an arbitrary direction within the x-y plane.  Since the value of $\tau_z$ is conserved by the Hamiltonian in the absence of interlayer tunneling, the Hamiltonian is invariant under rotations about pseudospin z-axis, so a non-zero expectation value of a polarization component in the x-y plane means that the U(1) symmetry has been spontaneously broken.  In the Hartree-Fock approximation, the ground state is a single filled Landau level for electrons with pseudospins aligned in the selected direction, so the ground state is often referred to as 
an {\em  interlayer-coherent integer quantized Hall state.}     (The state is also referred to as a quantum Hall ferromagnet.)\cite{ChakrabortyP1987,YoshiokaMG1989,fertig1989,MurphyEBPW1994,EisensteinM04,eisenstein2014}

The interlayer coherent state is found to have an energy gap, essentially the Hartree-Fock exchange energy, for any deviation in the total electron density away from  $\nu_{\rm{tot}}=1$. However, there is no energy gap for variations in the difference of the densities in the two layers, and this difference can be varied continuously within the coherent state.  Elementary excitations associated with variations in the relative density  involve  dynamically coupled  oscillations in the orientation of the in-plane component of $\sex{\vec{\tau}}$.   They are  Goldstone modes  resulting from the broken U(1) symmetry, and  they  have an energy that vanishes proportional to their wave vector $q$.      They may alternatively be described as magnons, in the ferromagnetic analogy.

An alternate way of describing the interlayer coherent state is to start from a state consisting of  a filled Landau layer in one layer and an empty Landau layer in the other. One may then add excitons, consisting of an electron in the empty layer bound to a hole in the filled layer.   Under appropriate circumstances,  when a finite density of excitons is present,    the ground state may be described as a Bose condensate of these excitons. Then the 
 the broken U(1) symmetry may be  described as a broken {\em phase symmetry} for the Bose condensate. 
 As would be expected for a Bose condensate,  the interlayer coherent state behaves like a superfluid for movement of the excitons. As neutral particles, excitons do not carry a net electric current, but an exciton current will carry a counterflow electrical current, which is equal and opposite in the two layers. The superfluid property means that in a state of constant current, there can be no gradient in the chemical potential for excitons, so any voltage gradients must be identical in the two layers.  Since the system is in a quantized Hall state for the net current, currents can flow without dissipation, and any net current will lead to a quantized Hall voltage in both layers.  
 
  During the 1990's and early 2000's,  properties of  the interlayer coherent state were extensively explored, theoretically and experimentally, in GaAs double-well systems. The most dramatic effects were manifest after techniques were developed to make separate electrical contacts to the two layers, In situations where tunneling between layers could be neglected, one could observe directly the phenomenon of counterflow current without a voltage drop and the closely related phenomenon of quantized Hall drag.\cite{EisensteinM04}   In  particular,  when a current $I$ is allowed to pass through one layer, while no current is allowed to flow in the second layer, a quantized Hall voltage of $I h/e^2$ is observed in {\it{both}}  layers, with no longitudinal voltage in either layer.

In the case where there is weak tunneling between the layers, the tunneling operator will break the U(1) symmetry of the Hamiltonian. If the applied magnetic field is perpendicular to the  layers, one may choose a gauge in which the tunneling operator is proportional to a particular combination of  $\tau_x$ and $\tau_y$, independent of position.  An arbitrarily small amount of tunneling will then align the pseudospin of the ground state so that its x-y component points in the favored  direction.   
Furthermore, since the value of $\tau_z$ is no longer conserved, its value must now be chosen to minimize the total energy. If the Hamiltonian is symmetric with respect to occupation of the two layers, the lowest energy state will have $\sex{\tau_z} = 0$. However, $\sex{\tau_z}$ will be different from zero, if there is a bias favoring one layer or the other.

If  interlayer tunneling is important, it is not possible to conduct a drag experiment in which current flows in the two layers are separately controlled.  However, one can apply a voltage difference between the two layers and measure the tunnel current between the layers.  In the interlayer coherent state, there will be a sharp increase in the tunneling current at very low  voltage difference between the layers, analogous to Josephson tunneling in a superconductor.  

In GaAs samples where the layers are close enough for the interlayer coherent state to occur, there will  typically be tunneling between layers that  cannot be neglected at $\nu_{\rm{tot}}=1$, if  the applied magnetic field is perpendicular to the sample.  However tunneling effects can be suppressed in a tilted magnetic field.   The in-plane field component then gives a momentum boost to electrons tunneling between the layers, which can destroy the coherent Josephson effect.\cite{eisenstein2014}  

Very recently, it has been possible to construct  Coulomb-coupled bilayer systems using two graphene sheets, separated by an insulating layer of about 2.5 nm of hBN. In this system, tunneling is  negligible even in the absence of an in-plane magnetic field, because of the large energy gap of hBN.  Results of various experiments on Coulomb-coupled graphene systems are discussed  in the chapter by Dean, Kim, Li and Young. 

\subsection {Transition to the Interlayer Coherent State}

On the theoretical side, there is much interest in what happens to the interlayer coherent state  in an ideal sample as the separation between layers is increased.
The most important variable determining the relative importance of interlayer and intralayer Coulomb interactions is the ratio between   the interlayer separation $d$ and the magnetic length $l_B$. Experimentally, it is difficult to vary $d$ continuously,  but one can vary the electron density and hence the value of $l_B$ at a given value of $\nu_{\rm{tot}}$.
Additional issues concern the effects of finite temperature, of imbalance between the layers, of disorder, and/or of weak tunneling between layers.

For two spin-polarized layers with no disorder and $\nu =1/2$ in each, one might  expect that for sufficiently large separations, the ground state would consist of two uncoupled layers, with a Fermi surface and gapless excitations in each layer. However, in analogy with the Cooper instability to superconductivity for a conventional Fermi surface, one should consider the possibility of pairing instabilities of the coupled composite fermion seas for arbitrarily weak interactions. Instabilities have been considered in several channels. The situation is complicated because of the long-range gauge fields and resulting infrared divergences at the Fermi-surfaces of the separated layers. Nevertheless, it is generally believed that there should be an energy gap of some kind in the ideal case for arbitrary separations of the layers,  though the gap may decrease very rapidly with increasing layer separation and may be completely unmeasurable for large separations.  What is not clear is what will be the symmetry of the resulting gapped state. One possibility is that the gapped state at large separations is topologically distinct from the interlayer coherent state, in which case there must be one or more sharp phase transitions between the ideal ground states as the separation is increased, Another possibility, however, is that the states are topologically identical, so that in principle there could be only a crossover, and no sharp transition at any finite separation, as the separation is increased. This possibility is not immediately obvious, as the interlayer coherent state is thought of as a condensate of excitons made from particles and holes with no gauge interactions, while, in the HLR picture, the separated layers are described by two Fermi surfaces of electron-like CFs, with two separate gauge fields.

A recent analysis by Sodemann {\it{et al.}} \cite{sodemannKWS19} has explained in some detail  how this can occur  from the point of view of both  the Son-Dirac and HLR approaches. 
Within the HLR  approach,   Sodemann {\it{et al.}} find that one can realize a state that is topologically equivalent to the interlayer coherent integer quantized state by coupling the two electron-like CF systems in a BCS-like pair state with $p_x + i p_y$  symmetry.  A trial wave function embodying this type of pairing for CFs was introduced  in 2008 by M\"oller et al., whose energy could be evaluated by Monte Carlo techniques and 
which could be compared with exact diagonalization studies for small systems.\cite{MollerSR08}  They found very good agreement for intermediate layer separations ($d/l_B \approx 1$) but poorer agreement at small separations. However, they were able to obtain good overlap with exact calculations at smaller separations by using a slightly more complicated wave function with the same symmetry as the pairing wave function.\cite{MollerSR09}  For large separations, the pairing amplitude becomes very small, and  trial wave function cannot be distinguished from two independent CF liquids in a system of limited size.   

An alternate  approach to describing the ground state at various layer separations would be to first make a particle-hole transformation in one of the two layers, and attach Chern-Simons flux to the holes in that layer. One can then create a state topologically equivalent to the interlayer coherent state by pairing the hole-like CFs and the electron-like CFs in a conventional $s$-wave BCS state. So far, however, there have been no numerical calculations using trial wave functions based on this construction.


The question of whether there is a actually a smooth crossover between the behavior of two separated layers and the interlayer coherent state, or whether there some type of phase transition separating different phases has been examined in numerical calculations of finite systems.\cite{MollerSR08,MollerSR09,MilovanovicDP15}  However, there does not seem to be a firm conclusion about this issue.

The situation at balanced layer filling may be contrasted with the situation at $\nu_{\rm{tot}}=1$ with unequal filling.  For an ideal isolated layer, there should be an infinite number of stable gapped states with filling factors of the Jain form $\nu = p/ (2p +1)$. As these states each have a finite energy gap, a pair of such layers  with Jain filling factors $\nu_1$ and $\nu_2 = 1 -\nu_1$  would  not be sensitive to very weak interactions between the layers, so they would remain as independent gapped states for sufficiently  large separations.  These separately gapped  states could not be connected, without a phase transition, to the interlayer coherent state, which has no energy gap for the difference in densities of  the layers. 

What happens if the density in each layer is intermediate between two neighboring Jain states?  For an isolated layer with short range interactions, it is expected that close to $\nu=1/2$, there will be  a series of first-order phase transitions between quantized states with successive integer values of $p$, which implies that for intermediate filling factors the system would phase separate into domains of Jain states with different densities. With unscreened Coulomb interactions, one  would  expect there to be no true phase separation; rather, the system would form an array of large finite domains, in a structure with a broken translational invariance whose period would be sensitive to the precise filling factor of layer. In any case, if the well-separated layers have unequal fillings with $\nu_{\rm{tot}}=1$, there must be at least one sharp phase transition as the separation is decreased, if the final state is a translationally-invariant interlayer coherent state. 

In experiments
on GaAs bilayer systems, it appears that there is a sharp phase transition, even in the case of balanced layer occupation,   as one increases the value of $d/ l_B$ through a critical value of order 1.8.\cite{SpielmanKEPW04}


In recent work, Eisenstein {\it{et al.}} \cite{eisenstein19} have studied the tunneling current between layers of a GaAs heterostructure in the regime of $d/l_B$ slightly larger than the critical value for formation of the interlayer-coherent quantized state at $\nu_{\rm{tot}} = 1$.  Measuring the tunnel current as a function of bias voltage, they found that the Coulomb pseudogap, which generally inhibits interlayer tunneling at low energies, was itself suppressed as the critical value of $d/l_B$ was approached.  
The results which could be interpreted as a reflection of interlayer electron-hole  correlations, which became stronger as the critical value of $d/l_B$ is approached. Studies of the dependence of the pseudogap on the density difference between the 
two layers found a larger suppression effect at filling-factor differences $\Delta \nu = \pm 0.08$ than at equal filling. This may reflect the curious fact, previously observed but not yet understood, that the critical value of $d/l_B$ for formation of the quantized interlayer coherent state is actually a local minimum at $\Delta \nu = 0$.\cite{SpielmanKEPW04} From a larger perspective, these experiments demonstrate the usefulness of tunneling measurements for investigating subtle correlations between Coulomb-coupled layers in the quantum Hall regime.

In general, a quantized Hall conductance is theoretically precise only in the limit of zero temperature.  Because charged excitations in a quantized Hall state have finite energy, there will be a non-zero density of such excitations at any finite temperature, and one does not expect to have a sharp phase transition as the temperature is increased. However, in the interlayer coherent state, the excitations which can destroy  exciton superfluidity are vortices in the Bose condensate, which have a logarithmically diverging energy at $T=0$.  Thus, one would expect a mathematically well defined-phase transition as the temperature is raised,  of the Kosterlitz-Thouless type, where counterflow superfluidity is lost.  

Experiments to study the temperature-driven  transition have been carried out in Coulomb-coupled graphene double layers.\cite{liu2019}   Interpretations are complicated by possible effects of disorder and because observation of the transition required finite measuring currents.  Nevertheless, the experiments have enabled one to track  behavior of the transition temperature as one varies the electron density, and hence  $d/l_B$.  It was found that the effective transition temperature increases with density for  small values of $d/l_B$, as one would expect for a Bose condensate of tightly bound excitons.  However around $d/l_B$ of order 0.8, as the excitons begin to overlap more strongly, the transition temperature starts to decreased with increasing density.  This is what  one might expect  qualitatively if one is entering a regime of  BCS-like pairing, as the Fermi level will rise with increasing density, and the ratio of the interlayer interactions to intralayer interactions will decrease. 

In the regime of relatively large $d/l_B$, the counterflow resistance was found to increase sharply near the critical temperature, with a dependence on temperature and measuring current similar to what one would expect for a Kosterlitz-Thouless transition. On the other hand, for smaller values of $d/l_B$, the temperature dependence of the counterflow resistance  was better fit by a simple Arrhenius law, down to the lowest temperatures where it could be observed.  Arrhenius behavior has been also observed for GaAs double layers. This has been attributed to effects of disorder, which might lead to a finite density of vortices even at $T=0$, which then might be unpinned and contribute to transport at finite temperatures.\cite{EisensteinM04} However, a full explanation for these observations has not been developed.

\section{Conclusions}
 
Although problems related to quantum Hall phenomena at, or in the vicinity of  a half full Landau level have been actively studied for nearly three decades, the subject remains active and exciting, due to new experimental observations and to new theoretical developments.  In this chapter, we have reviewed the current status of the field, with emphasis on new developments and unanswered questions. A major focus of the chapter has been issues surrounding systems of spin-polarized electrons  in the lowest Landau level, as in GaAs narrow quantum wells, where no quantized Hall plateau is observed at filling $\nu = 1/2$. However, we have also discussed, briefly, situations where a quantized Hall plateau is observed, such as at $\nu = 5/2$ and $\nu=7/2$ in GaAs structures.  Finally,  we have discussed phenomena in double-well systems where the half full condition is achieved when the total filing factor is $\nu_{\rm{tot}}=1$.

We find that there remain many unanswered questions in the field.
 
\section* {Acknowledgments}

Over the years, I have benefited greatly from discussions with many colleagues, too numerous to name, on  the various topics covered in this chapter.  However, I would like to thank particularly Chong Wang, Ady Stern, and Jainendra Jain for their helpful comments on preliminary versions of the manuscript.


\bibliographystyle{ws-rv-van}
\bibliography{half-full-chapter}

\end{document}